\documentclass[twocol]{ametsocV6.1}
\usepackage{enumitem}

\title{A fluctuation-dissipation theorem perspective on radiative\\ responses to temperature perturbations}

\authors{Fabrizio Falasca,\aff{a}\correspondingauthor{Fabrizio Falasca, fabri.falasca@nyu.edu} 
Aurora Basinski-Ferris,\aff{a} 
Laure Zanna,\aff{a} 
Ming Zhao\aff{b}
}

\affiliation{\aff{a}{Courant Institute of Mathematical Sciences, New York University, New York, NY, USA }\\
\aff{b}{NOAA Geophysical Fluid Dynamics Laboratory, Princeton, New Jersey, USA}\\
}

\abstract{
Radiative forcing drives warming in the Earth system, leading to changes in sea surface temperatures (SSTs) and associated radiative feedbacks. The link between changes in the top-of-the-atmosphere (TOA) net radiative flux and SST patterns, known as the “pattern effect”, is typically diagnosed by studying the response of atmosphere-only models to SST perturbations. In this work, we diagnose the pattern effect through response theory, by performing idealized warming perturbation experiments from unperturbed data alone. First, by studying the response at short time scales, where the response is dominated by atmospheric variability, we recover results that agree with the literature. Second, by extending the framework to longer time scales, we capture coupled interactions between the slow ocean component and the atmosphere, yielding a novel “sensitivity map’’ quantifying the response of the net radiative flux to SST perturbations in the coupled system. Here, feedbacks are captured by a spatiotemporal response operator, rather than time-independent maps as in traditional studies. Both formulations skillfully reconstruct changes in externally forced simulations and provide practical strategies for climate studies. The key distinction lies in their perspectives on climate feedbacks. The first formulation, closely aligned with prediction tasks, follows the traditional view in which slow variables, such as SSTs, exert a one-way influence on fast variables. The second formulation broadens this perspective by incorporating spatiotemporal interactions across state variables. This alternative approach explores how localized SST perturbations can alter the coupled dynamics, leading to temperature changes in remote areas and further impacting the radiative fluxes at later times.}

\begin{document}
\maketitle

%
%
%
%
%

%

\maketitle

\section{Introduction}

The response of Earth's surface temperature to changes in atmospheric $\textrm{CO}_2$ concentration is at the heart of climate change science and yet remains uncertain \citep[e.g.,][]{IPCC_2022_WGIII_Ch_1,zelinka2020causes,meehletal2020}. For example, the equilibrium global mean surface temperature change to a doubling of $\textrm{CO}_2$, i.e., ``equilibrium climate sensitivity'', ranges between $1.8-5.6^o$C in the latest generation of climate models, an even larger uncertainty than the previous Coupled Model Intercomparison Project (CMIP) \citep{meehletal2020}. A common framework to understand the global mean climate response to external forcing is through the linear global energy balance, in which we decompose the net heat flux ($N$) at the top of the atmosphere as a radiative forcing ($Q$) and a radiative response of the system ($H$) as $N = Q - H$ \citep{Gregory2004ASensitivity}. The radiative response can be expressed as $H \approx \lambda \Delta T$, where $\lambda$ $[W/(m^2 K)]$ is a climate feedback parameter and $\Delta T$ $[K]$, the global mean temperature change \citep{Gregory2004ASensitivity}. 
The magnitude of the climate feedback parameter $\lambda$ is a large contributor to the uncertainty of climate sensitivity in models \citep{chenal2022observational,roe2011sensitive}, primarily due to the poor representation of clouds \citep[e.g.,][]{zelinka2020causes}.
\\\\ \noindent
The climate feedback parameter ($\lambda$) is often approximated as constant; however, it has been demonstrated in coupled climate models that $\lambda$ evolves in time, even under a time-independent CO$_2$ forcing \citep[e.g.,][]{murphy1995transient,senior2000time,williams2008time,andrews2015dependence,Meyssignac}. It is generally hypothesized that the temporal evolution of $\lambda$ is due to the evolution of the spatial pattern of surface temperature, which can initiate different climate feedbacks over time \citep{armour2013time,Zhao22investigation}, referred to as the ``pattern effect'' \citep{stevens2016prospects}. In other words, given the same global mean temperature change, different spatial patterns of sea surface temperature change can lead to very different radiative feedbacks. Given the importance of $\lambda$ for constraining climate sensitivity, the pattern effect has been a source of continuous investigation in recent years \citep{zhouetal2017,Zhang2023UsingCM4,Dong2020IntermodelModels,Dong2019AttributingPacific, alessi2023surface,bloch2020spatial,kang2023recent, blochjohnsonetal2024}.\\

The current approach for diagnosing the pattern effect relies on Green's function experiments \citep{barsugli2002global}, by leveraging atmosphere-only models to diagnose the response of the top of the atmosphere (TOA) radiative fluxes to perturbations in the sea surface temperature field \citep{zhouetal2017,Zhang2023UsingCM4,Dong2020IntermodelModels,Dong2019AttributingPacific, alessi2023surface}. Despite some variations, current studies tend to have the following features in common, as summarized in \cite{blochjohnsonetal2024}: (i) the assumption that global mean TOA radiative response is linearly related to variations in the sea surface temperature (SST) spatial pattern, (ii) utilizing an atmosphere-only model to diagnose the atmosphere's sensitivity to SST boundary conditions, and (iii) using relatively large perturbations (e.g., $1-4K$) in the SST field when constructing the Green's functions. Large perturbations are beneficial for detecting a clear response with a shorter integration time but also lead to additional problems as large perturbations can result in nonlinear responses \citep{williams2023circus}, invalidating assumption (i) stated above. In addition to the Green's function approach, there have been efforts to estimate a response operator from existing simulation output to avoid the computational cost of numerical perturbation experiments.
In particular, \cite{zhouetal2017,bloch2020spatial,kang2023recent} have all used forms of linear regression to estimate the pattern effect and explore sensitivities from internal variability. Regression approaches successfully eliminate the computational expenses of previous Green’s function methods, but may not be optimally designed to infer spatiotemporal-dependent causal links among climate fields.\\

In this work, we present a method for diagnosing the pattern effect based on the Fluctuation-Dissipation Relation, also often referred to as Fluctuation-Dissipation Theorem (FDT) \citep[see, e.g.,][]{MajdaBook}, and response theory \citep{Marconi}. The Fluctuation-Response formalism provides a strategy to compute the ensemble average response of a physical system to \textit{external} perturbations, solely given correlation functions of the \textit{unperturbed} system \citep{Kraichnan,Kubo1966,MajdaBook,Marconi}. The Earth's climate is a multiscale, complex dynamical system for which the application of the FDT formalism is only possible by focusing on the proper observables \citep{GritsunValerio} and on a narrow range of spatiotemporal scales \citep{MajdaBook,Provenz}. The protocol proposed here, building on the framework recently presented in \cite{FabCausal}, leverages the FDT formalism together with dimensionality reduction techniques to infer a response operator in a coarse-grained representation of the climate system. The success of linear response theory for climate data is closely tied to coarse-graining methods \citep{Colangeli2012}. In the case of spatiotemporal dynamical systems (such as climate), coarse-graining procedures refer to (i) averaging over large spatial regions (or choosing a few relevant modes), (ii) selecting a limited range of temporal scales, and (iii) considering a limited set of variables. These coarse-graining steps need to be carefully considered in applications of FDT in climate and they will be detailed throughout this paper. The response operator, together with the appropriate convolution formulas, allows us to study the causal relation between the SST and the TOA net radiative flux fields across multiple spatial and temporal scales.\\

The method presented in this work bridges the gap between Green’s function approach \citep{blochjohnsonetal2024} and previous statistical methods \citep{zhouetal2017,bloch2020spatial}. Like Green’s function techniques derived from atmospheric models \citep{Zhang2023UsingCM4}, our framework infers causal linkages among climate fields within the paradigm of responses to external perturbations \citep{Baldovin}. At the same time, it remains data-driven, akin to prior statistical approaches, offering a computationally fast, causal protocol to diagnose the pattern effect directly from data.\\ 

We apply the proposed framework to the GFDL-CM4 model \citep{cm4}, demonstrating its relevance to the literature on the pattern effect in several ways. First, by focusing on responses at the shortest time scales, where the atmospheric dynamics dominates the response to SST perturbations, we recover results that align with the existing literature. This strategy is justified as the response of TOA fluxes to changes in SST is fast (e.g., $\sim20$ days in tropical large-scale convective overturning motion). We refer to this formulation as ``atmospheric-only'' formulation. In a second step, we incorporate responses over longer time scales and refer to this as the ``atmosphere-ocean coupled'', or ``coupled'', formulation. Here, the slow ocean response to SST perturbations becomes significant, and the net flux at the TOA can be further modulated by slow changes driven by the system's coupled dynamics reacting to the initial perturbation patterns. This extended analysis yields a novel sensitivity map, characterized by negative sensitivity across the tropical Pacific, in contrast to the traditional view of a negative-positive dipole in the basin. Both formulations of the protocol demonstrate good predictive skill in capturing changes in radiative flux at the TOA across two independent forced experiments. In Section \ref{sec:Feedbacks}, we argue that in the specific case of pattern effect, where TOA responds very fast to a change in the temperature, the two reconstructions should converge in the limit of infinite data and complete knowledge of all climate variables. Both approaches are formally ``correct''; their primary distinction lies in their perspective on climate feedbacks. The first formulation adopts the traditional view, closely linked to prediction tasks, treating feedbacks as quasi-instantaneous, with one-way directionality from slow variables (e.g., SST) to fast ones (e.g., TOA radiative flux). In contrast, the second formulation introduces a novel perspective where initial perturbations at time $t$ drive cascades of responses in the coupled climate dynamics at $t + \tau$, which can further amplify or dampen the radiative balance. Feedbacks, in this case, are encoded in a spatially and temporally dependent response operator rather than in static maps. This distinction builds on the framework proposed by \cite{Lucarini2018}, also in the context of response theory, and formalizes ideas and considerations previously discussed in the climate literature by \cite{PascaleSW}. From an information theory perspective, this view of feedbacks is closely linked to the idea of ``flow information'' as first shown in \cite{Ay}. As further detailed in Sections \ref{sec:Feedbacks}, \ref{sec:conclusion} and Appendix A, this ``coupled'' formulation also introduces new challenges, as it requires accounting for the full complexity of the climate system's response. In the coupled system, focusing on only two variables represents a more significant simplification than in the atmosphere-only system.\\

Importantly, in both formulations, we eliminate the need to prescribe large perturbations to the system to construct the Green's function operator. This ensures the validity of the linearity assumption when studying the response of TOA fluxes to SST perturbations. The FDT response operator is derived in the limit of infinitesimally small perturbations, where the linearity assumption holds \citep{MajdaBook,Marconi}. Evaluation tests presented in this paper demonstrate that, in the specific context of the pattern effect and with the proposed coarse-graining procedure, the FDT formalism can also provide accurate predictions for global mean TOA response to finite-amplitude perturbations in the SST pattern.\\

In what follows, we describe the proposed methodology and data preprocessing in Sections \ref{sec:methods} and \ref{sec:data}. We present the general, data-driven protocol to perform perturbation experiments in a multivariate climate system in Section \ref{sec:protocol}. The ``atmospheric-only'' and ``coupled'' formulations are presented in Sections \ref{sec:Atmos-Only-Protocol} and \ref{sec:Coupled-Protocol} respectively. The main differences across approaches, mainly related to a different perspective on climate feedbacks are further outlined in Section \ref{sec:Feedbacks}. Conclusions and future work are given in Section \ref{sec:conclusion}. Useful considerations for the correct application of FDT to realistic climate data, including limitations and caveats, are further outlined in detail in Appendix A and are especially relevant for future applications of FDT.

\section{Proposed theoretical framework} \label{sec:methods}

We present our proposed framework, which focuses on computing a response operator from the unforced fluctuations of the climate system. In practice, to meet the necessary assumptions in linear response theory, we compute the response operator for a coarse-grained representation of the relevant state variables by focusing on large-scale spatial averages and a limited range of time scales. We detail the dimensionality reduction technique in Section \ref{sec:methods}\ref{sec:dim_red} and the data preprocessing in Section \ref{sec:data}. Further details on the necessary assumptions and implementation choices are discussed in Appendix A.

\subsection{Linear response theory and the Fluctuation-Dissipation Theorem} \label{sec:Lin_Resp_T}

Consider a dynamical system with state vector $\mathbf{x}(t) = [x_{1}(t),x_{2}(t),...,x_{N}(t)]$. Here, $N$ is the dimensionality of the system and $t$ is a time index. We study the system from the perspective of stochastic dynamics, interpreting $\mathbf{x}(t)$ as a set of $N$ coarse-grained variables. The integrated effects of the unresolved degrees of freedom on the resolved ones are modeled as suitable noise terms \citep{Hasselmann,Penland89,MTV1,MTV2,LucariniChekroun}. We perturb the system by applying a small, time-dependent perturbation $\delta x_j(t) = \Delta x_j \delta f_j(t)$ directly to the $j$-th degree of freedom $x_j$, i.e. $x_j \rightarrow x_j + \delta x_j(t)$. The leading order of the response in the $k$-th variable can be written as: 
\begin{equation}
\begin{split}
\delta \langle x_k(t) \rangle &= \langle x_k(t) \rangle_\text{p} - \langle x_k(t) \rangle \\ 
&= \sum_j \int_{0}^{t} R_{k,j}(\tau) \delta x_j(t-\tau) d\tau,
\end{split}
\label{eq:response_convolution}
\end{equation}
where $\langle \cdot \rangle$ and $\langle \cdot \rangle_\text{p}$ indicate ensemble averages of the system before and after applying the perturbation. The response operator $R_{k,j}(t)$ is defined as the average response of the variable $x_{k}(t)$ to an \textit{impulse} perturbation $\Delta x_j \delta(t)$ applied to the variable $x_{j}(0)$, where $\delta(t)$ denotes the Dirac delta function \citep{Risken,Baldovin2022}.  If the perturbation $\delta x_j(t)$ is instead represented as a Heaviside step function, $\delta x_j(t) = \delta x_j \Theta(t)$, the response simplifies to:
\begin{equation}
\begin{split}
\delta \langle x_k(t) \rangle = \sum_j \delta x_j \int_{0}^{t} R_{k,j}(\tau) d\tau.
\end{split}
\label{eq:response_step}
\end{equation}
The primary challenge in quantifying responses to perturbations, as described in Eqs. \eqref{eq:response_convolution} and \eqref{eq:response_step}, is to infer the response operator in terms of the \textit{unperturbed} system. The Fluctuation-Dissipation Theorem (FDT) addresses this challenge by establishing a causal link between the variability of the \textit{unperturbed}, stationary system and its response to \textit{external} perturbations \citep[e.g.,][]{Marconi,Hairer2010}. Verifying the existence of such a link, for specific cases or variables of interest, is highly relevant for climate dynamics, whether to study the climate response to external forcing \citep{Leith,MajdaBook} or to analyze the spatio-temporal interactions among different climate variables \citep{Lucarini2018,FabCausal}.

\paragraph{FDT: general formulation.} Given a sufficiently smooth and non-vanishing invariant probability distribution, $\rho(\mathbf{x})$, of the stochastic system $\mathbf{x}(t)$, the following result holds:
\begin{equation}
R_{k,j}(t) = \lim_{\delta x_{j}(0)\to0} \frac{\delta \langle x_{k}(t) \rangle}{\delta x_{j}(0)} = - \Big\langle x_{k}(t) \frac{\partial \ln \rho(\mathbf{x})}{\partial x_{j}} \Big|_{\mathbf{x}(0)} \Big\rangle .
\label{eq:response_general} 
\end{equation}
This formulation represents a general form of the FDT \citep{FALCIONI}. Eq. \eqref{eq:response_general} enables us to evaluate the response of a stochastic dynamical system to infinitesimal impulse perturbations using only the unperturbed dynamics of the system. We note that by focusing on stochastic systems, we circumvent the technical difficulties associated with deterministic dissipative dynamical systems, where the invariant measure is singular \citep[see][]{RUELLE1998220,RuelleReview,Colangeli2012}.

\paragraph{FDT and causality.} Recent literature including \cite{Aurell,Lucarini2018,Baldovin} has pointed out the connection between the Fluctuation-Response formalism and the role of causality in physical systems based on the notion of intervention \citep{PearlBook,IsmaelNew}. The main idea is that, in physical experiments, cause-effect relations are inferred by probing the system and examining its response. Specifically, the link $x_j(t) \rightarrow x_k(t+\tau)$ is inferred by studying how an \textit{external} perturbation at variable $x_j(t)$ propagates along the system, inducing \textit{on average} a change in variable $x_k(t+\tau)$. In the case of small perturbations, the Fluctuation-Dissipation Theorem shown in Eq. \eqref{eq:response_general} allows us to do so in a straightforward way, by inferring what the response would have been if we had perturbed the system\footnote{We stress the important difference between methodologies for ``causal discovery'', aimed at reconstructing a causal graph from time series, and the more general task of causal inference. As an example, given three variables $\{x,y,z\}$ and a causal graph such as $x \rightarrow y \rightarrow z$, the objective of a causal discovery method will be to discover the graph itself. Causal inference requires us to go one step further and study the effect of interventions on the graph; the variability of both variables $x$ and $y$ will cause the variability of $z$. See \cite{Ay} and \cite{RungeCausalEffects} for details.}. We refer the reader to \cite{Aurell} and \cite{Baldovin} for further details and to \cite{Provenz,Cecconi,FabCausal,ResponseScore} for some examples of applications. 
\paragraph{Quasi-Gaussian approximation.} The main practical issue with the formulation given by Eq. \eqref{eq:response_general} is that the functional form of $\rho(\mathbf{x})$  is not known a priori and is highly nontrivial in high-dimensional systems. While recent promising results in the estimation of $\nabla_\mathbf{x}\ln\rho(\mathbf{x})$ using generative modeling \citep{ResponseScore} have emerged, generally, the strategy has been to approximate $\rho(\mathbf{x})$ as a Gaussian distribution \citep{Leith}. In the case of Gaussian distributions, Eq. \eqref{eq:response_general} reduces to:
\begin{equation}
\mathbf{R}(t) = \mathbf{C}(t)\mathbf{C}(0)^{-1},
\label{eq:response_linear} 
\end{equation}
with the covariance function $C_{i,j}(t) = \langle x_{i}(\tau+t) x_{j}(\tau) \rangle$ ($x_{i}$ is assumed to be zero mean). Eq. \eqref{eq:response_linear} is valid for linear systems and has been referred to as the ``quasi-Gaussian approximation'' \citep{MajdaBook}. The form of FDT shown in Eq. \eqref{eq:response_linear} motivated many studies in climate, including \cite{GRITSUN,Ring,Majda2010,Pedram1,Pedram2,Christensen,HeldFDT}, which focused on the implications or limitations of this formalism. It has been shown empirically that the quasi-Gaussian approximation has high skill for predicting the response in the ensemble mean in non-Gaussian regimes \citep{GRITSUN,GERSHGORIN20101741,Baldovin} and has some skill for the response in variance \citep{GritsunMajda,Majda2010}. In fact, while the climate system is a high-dimensional, chaotic dynamical system, the probability density of many coarse-grained variables is often smooth and nearly Gaussian \citep{MajdaStructuralStability}. The quasi-Gaussian approximation is then relevant in climate studies after the appropriate spatial and temporal coarse-graining.

\subsection{Numerics and statistics} \label{sec:numerics_statistics}

The estimation of the response operator $\mathbf{R}(t) = \mathbf{C}(t)\mathbf{C}(0)^{-1}$ is limited by the data sample size and by the \textit{effective} dimensionality of the dynamics \citep{GRITSUN,Martynov}. These limitations lead to two primary challenges: (i) the inferred covariance matrix $\mathbf{C}(0)$ is often ill-conditioned, leading to significant errors in the computation of its inverse; (ii) $\mathbf{R}(t)$ is contaminated by spurious terms due to the interplay of limited sample size and strong autocorrelations of the underlying time series. To address these issues, we employ both a regularization procedure when computing the covariance matrix and an estimation of the confidence bounds introduced in \cite{FabCausal}. We stress that the effectiveness of these tools depends on the problem at hand, including factors such as the amount of available data and the dimensionality of the system.

\paragraph{Regularization strategy.} 
Computing the covariance matrix, $\mathbf{C}(0)$, after reducing the dimensionality of the system, helps to lower the condition number. However, the condition number may remain high enough that a regularization step is warranted. In our case, we compute the covariance matrices in a low-dimensional space (see Section \ref{sec:methods}\ref{sec:dim_red}) and add a Tikhonov regularization \citep{Hansen} as follows:
\begin{equation}
\mathbf{C}_{r}(0) = \mathbf{C}(0) + \lambda \mathbf{I},
\label{eq:Tikhonov} 
\end{equation}
where $\mathbf{C}_{r}(0)$ represents the regularized matrix, $\lambda$ is a parameter and $\mathbf{I}$ is the identity matrix. In practice, there is a balance between lowering the condition number of the covariance matrix and retaining important elements of the system's dynamics; if the choice of $\lambda$ is too large, it can obscure significant correlations. Here, we regularize the data as follows: (i) we compute the maximum eigenvalue $\lambda_{\text{max}}$ of the matrix $\mathbf{C}(0)$; (ii) we choose a regularization parameter $\lambda = 10^{-2} \lambda_{\text{max}}$, resulting in covariance matrices $\mathbf{C}(0)$ with condition numbers $\sim 100$.

\paragraph{Confidence bounds.} The theoretical tools presented in Section \ref{sec:methods}\ref{sec:Lin_Resp_T} require computing covariances through ensemble averages, which is impossible in climate applications due to the limitations of a single realization. The common way to overcome this is through the assumption of ergodicity, therefore replacing ensemble averages with temporal averages \citep{nonEqStatMech}. Covariance matrices are then computed using temporal averages $C_{i,j}(t) = \overline{x_{i}(\tau+t) x_{j}(\tau)}$, i.e. $\overline{f}$ being the temporal average of $f$. This method will result in spurious responses $\mathbf{R}(t)$, because of (i) finite sample size (i.e., the length $T$ of the trajectory is finite) and (ii) large autocorrelations of the time series $x_i(t)$. In order to identify spurious results of the response operator, we adopt the confidence bounds proposed in \cite{FabCausal}. Under the null hypothesis of a multivariate red noise process, a choice relevant for climate data \citep{AllenSmith,Dijkstra2013}, it is possible to derive the null distribution of the response operator $\hat{\mathbf{R}}(t) = \hat{\mathbf{C}}(t)\hat{\mathbf{C}}(0)^{-1}$ with the following expected value and variance:
\begin{equation}
\begin{split}
\mathbb{E}[\hat{R}_{k,j}(t)] &= \phi_{k}^{t} \delta_{k,j};\\
\mathbb{V}\textrm{ar}[\hat{R}_{k,j}(t)] &= \frac{\sigma^{2}_k}{\sigma^{2}_j} \Big[ \frac{\phi_k^{2t}-1}{T}+\frac{2}{T} \Big{(}\frac{1-\phi_k^{t}\phi_j^{t}}{1-\phi_k\phi_j} \Big{)}\\ 
&- \frac{2\phi_{k}^{t}}{T} \Big{(} \phi_{k} \frac{\phi_{j}^{t}-\phi_{k}^{t}}{\phi_{j}-\phi_{k}} \Big{)} \Big] ~ .
\end{split}
\label{eq:expectation_and_Variance_R}
\end{equation}
where $\phi_{i}$, $T$ and $\sigma^2_i$ respectively represent the inferred autocorrelation, the length $T$ of each time series $x_i(t)$ and their (regularized) variances $[\mathbf{C}_r(0)]_{i,i}$. Here, $\delta_{k,j}$ is the Kronecker delta. The symbol $\hat{f}$ specifies statistics $f$ of the null model rather than of the original system. Finally, in the case $\phi_{k} = \phi_{j}$ we substitute the term $\phi_{k}\frac{\phi_{j}^{t}-\phi_{k}^{t}}{\phi_{j}-\phi_{k}}$ with $\phi_{k}^{t} t$.\footnote{We do so as $\lim_{\phi_{j}\to\phi_{k}} \phi_{k}\frac{\phi_{j}^{t}-\phi_{k}^{t}}{\phi_{j}-\phi_{k}} = \phi_{k}^{t} \tau$.} We refer to \cite{FabCausal} for details on the derivation of Eq. \ref{eq:expectation_and_Variance_R}. In this paper, linear response formulas as Eq. \eqref{eq:response_convolution} are computed after neglecting insignificant terms in the response operator. This work will focus on the $\pm 1 \sigma$ confidence level.

\subsection{Dimensionality reduction} \label{sec:dim_red}

The formulas proposed in the previous section cannot be applied to the original high-dimensional system. First, as outlined before, the covariance matrix $\mathbf{C}(0)$ becomes ill-conditioned without dimensionality reduction due to high correlations between neighboring time series (e.g. $x_i(t)$ and $x_{i+1}(t)$) in the original data. Second, focusing on large scale averages is an important step of the coarse graining procedure, which leads to smooth and Gaussian-like probability densities \citep{MajdaStructuralStability,Prashant}, thereby justifying the application of the theory from earlier Sections. Moreover, since the climate system resides on a low-dimensional attractor, we aim (at least in theory) to study the system in its \textit{effective} dimensional space \citep{GhilLucarini}.  This approach moves away from a grid-based representation of the system and instead emphasizes resolution-independent modes or patterns as fundamental components of the framework \citep{Predrag, Dubrulle}.\\ 

The method of spatial coarse-graining alone can have a large impact on results \citep[e.g.,][]{CROMMELIN,HeldFDT,Pedram2}. Climate studies using FDT overwhelmingly focused on Empirical Orthogonal Functions (EOFs) as the dimensionality reduction technique. However, \cite{HeldFDT} and \cite{Pedram2}  noted that the EOF step alone can be a major source of errors in FDT applications. Here, we are going to focus on a method recently proposed in \cite{FabCausal}. The method allows us to partition a climate field into a few regionally constrained patterns of highly correlated time series, leading to a large reduction in the number of degrees of freedom and, therefore, more robust and interpretable inference. More formally, consider a spatiotemporal field saved as a data matrix $\mathbf{x} \in \mathbb{R}^{N,T}$. $N$ is the number of grid points and $T$ is the length of each time series. For example, $\mathbf{x}$ could be the sea surface temperature field. The dimensionality reduction proposed in \cite{FabCausal} offers a simple strategy to partition this $N$ dimensional field in terms of $n$, non-overlapping patterns $c_1, c_2, c_3, ..., c_n$, with $n \ll N$. The methodology utilizes a community detection algorithm, Infomap \citep{MapEq}. Each $c_j$ represents a two-dimensional region defined as a \textit{regionally constrained} set of time series with large average pairwise correlation, which we will refer to as a pattern or region interchangeably. Finally, to each region $c_j$, we associate a time series defined as the integrated anomaly inside, i.e. $X(c_{j},t) = \sum_{i \in c_{j}} x_{i}(t) \cos(\theta_{i})$, where $\theta_{i}$ represents the latitude at grid point $i$ and $\cos(\theta_{i})$ is a latitudinal scaling. To summarize, given a spatiotemporal field saved as a data matrix $\mathbf{x} \in \mathbb{R}^{N,T}$, the proposed framework allows us to define a new field $\mathbf{X} \in \mathbb{R}^{n,T}$, with $n \ll N$. See Appendix B and \cite{FabCausal} for additional details.

\section{Data and preprocessing} \label{sec:data}

We focus on the state-of-the-art coupled climate model GFDL-CM4 \citep{cm4}, used in recent studies on the pattern effect \citep[e.g.,][]{Zhang2023UsingCM4}. In addition, it offers data from a long control run, necessary for trustworthy computations of the response operator in Eq. \ref{eq:response_linear}. The ocean component is the MOM6 ocean model \citep{OM4} with a horizontal grid spacing of 0.25$^\circ$ and 75 vertical layers. The atmospheric component is the AM4 model \citep{AM4a,AM4b} with a horizontal grid spacing of roughly 100 km and 33 vertical layers. There is additionally a land component (LM4) and a sea-ice component (SIS2). Our protocol, as detailed in Section \ref{sec:protocol}, can be applied to any model, although we focus on GFDL-CM4 here.\\

We consider three simulations of CM4: a pre-industrial control (piControl), and two idealized scenarios of $\text{CO}_2$ increase, namely $1$pctCO$_2$ and $4\times$CO$_2$. The piControl simulation is a 650-year-long, stationary run with constant $\text{CO}_2$ forcing at the preindustrial level. The $1$pctCO$_2$ and $4\times$CO$_2$ are idealized experiments simulating the climate system under a 1\% $\text{CO}_2$ increase per year and an abrupt increase of 4 times $\text{CO}_2$ concentration respectively. Both the $1$pctCO$_2$ and $4\times$CO$_2$ experiments start from the preindustrial $\text{CO}_2$ concentration and are run for 150 years. The linear response operator $\mathbf{R}(t)$, shown in Eq. \eqref{eq:response_linear}, is constructed using data from the piControl run. The forced experiments are used to test the method's performance. We consider two variables: sea surface temperature field (SST) and the global mean net radiative flux at the top of the atmosphere (TOA). We focus on the global mean TOA, i.e. one time series, rather than the full spatial field, as the spatiotemporal dynamics of TOA does not show large scale patterns of variability. In other words, it is a fast variable with respect to the slow SST variability and the main signal we are interested in is the global average. Note that this is a simplification compared to the current literature focusing on the whole TOA field. The TOA flux, hereafter referred to simply as ``TOA'', is computed as the (incoming shortwave) - (reflected shortwave) - (upward radiative longwave) fluxes \footnote{Each component of the TOA fluxes is referred in the CMIP6 \cite{cmip6} catalog as follows: incident shortwave: ``rsdt''; reflected shortwave: ``rsut''; the upward radiative longwave flux: ``rlut''.}, with all fluxes computed at the top of the atmosphere. The SST fields in each model experiment are remapped to $2.5^\circ$ by $2^\circ$ resolution; the original temporal resolution is $1$ day, but our analysis will focus on monthly and 6-month averages, therefore excluding fast variability at the daily temporal scale and focusing on the slow dynamics (see also Appendix A).

\subsection{Data preprocessing for the piControl run} \label{sec:preprocess_piControl}
We now consider the piControl run and perform four steps:
\begin{enumerate}[label=(\roman*)]
    \item Remove the first 50 years of the 650-year-long time series, given a short transient trend in the first few decades.
    \item Compute and store the climatology of each time series $y(t)$, calculated as the mean across all time steps $\mu = \overline{y(t)}$.
    \item Remove the periodic signal given by the seasonal cycle, therefore focusing on the stationary internal variability of the system. Specifically, we remove the average value of each month from the data (e.g., from each January, we remove the average across all Januaries, from each February, we remove the average across all Februaries, etc.). 
    \item High-pass filter the data with a cut-off frequency of $10^{-1}$ years. This allows us to remove (a) low frequency (e.g., multidecadal) oscillations, which are only sampled a few times even in a 600-year long experiment and (b) any additional slow drift that is present in the piControl run (for example, the Southern Ocean SST shows a slow drift in the control CM4 simulation). This choice tacitly assumes that the response of the net radiative flux to SST perturbation experiments emerges inside a 10-year window. We stress that the highpass filtering step is performed only in the control run.
\end{enumerate}
After this preprocessing, the resultant time series $y(t)$ have zero mean due to step (iii) and approximately meet the quasi-Gaussian assumption, largely due to the high-pass filtering in step (iv) and by considering monthly averaging (see Appendix C).

\subsection{Data preprocessing for forced experiments} \label{sec:preprocess_forced}
We now consider the 150-year long 1pctCO$_2$ and $4\times$CO$_2$ runs. As in the section above, we describe the preprocessing steps for a time series $y^{f}(t)$ encoding either the variability of SST at a specific location or the global mean net flux at the TOA. Here $f$ stands for ``forced''. We preprocess the forced data in the following way:
\begin{enumerate}[label=(\roman*)]
    \item Compute and store the climatology of each time series $y_i^f(t)$, calculated as the mean across all time steps  $\mu^f = \overline{y^f(t)}$.
    \item Calculate anomalies relative to the seasonal cycle by removing the average value of each month from the data (e.g., from each January, we remove the average value across all Januaries, from each February we remove the average across all Februaries etc.).
    \item Add the difference in means between the forced and control runs to retain the mean state difference despite removing seasonality; i.e., update $y^f(t)\leftarrow y^f(t)+\mu^f-\mu$.
\end{enumerate}

In the forced experiments, we remove the contribution to the radiative forcing coming from an increase in $\text{CO}_2$ concentration alone, allowing us to isolate and study the TOA radiative feedback to changes in SSTs. The global mean change in radiative forcing driven by changes in $\text{CO}_2$ alone scales approximately logarithmically with its concentration \citep{Pierrehumbert,Romps}. As is standard, we remove a constant of $8 Wm^{-2}$ from the TOA flux in the $4\times$CO$_2$ experiment \citep{Romps, Zhao22investigation}. We then remove a time-dependent correction of the form $\alpha \log [C_t/C_0]$ in the 1pctCO$_2$ run, where $C_0$ and $C_t$ are the concentration of $\text{CO}_2$ in the control run and forced experiment, respectively. We 
fit the $\alpha$ parameter by the change of $\sim 8 Wm^{-2}$ in the $4\times$CO$_2$ simulation. Thus, we have an additional step: 
\begin{enumerate}[resume,label=(\roman*)]
    \item Remove the contribution to radiative forcing coming from the $\text{CO}_2$ concentration alone.
\end{enumerate}

\paragraph*{Main assumptions and limitations.} The underlying assumption behind the preprocessing procedure is that the SST field and global mean TOA variables, coarse-grained in space and time, are the ``proper'' variables to study the pattern effect. The integrated effect of processes active at (i) small spatial scales and (ii) fast time scales will be considered as noise \citep{PENLAND1996534}. These are large simplifications, leading us to consider only two variables. As we will stress in the paper, this represents a more significant simplification in the case of the ``coupled'' formulation of the protocol, i.e. diagnosing how perturbations propagate throughout the system across spatial and temporal scales, compared to the more traditional ``atmosphere-only'' formulation. The limitations and strengths of these choices will be further discussed in the conclusions and in detail in Appendix A. Future studies will focus on adding more variables and different coarse-graining methods, but we believe this study offers a valuable starting point for future perturbation experiments in the coupled system.\\

\section{Practical implementation of the proposed framework} \label{sec:protocol}

We now present the steps towards the practical implementation of the theoretical framework in Section \ref{sec:methods}. The main steps can be summarized by three main points: given the original, high-dimensional fields, (i) project the dynamics in a lower-dimensional representation; (ii) compute the response formulas in the low-dimensional space; (iii) project the results back to the original, high-dimensional space.\\

As outlined in Section \ref{sec:data}, we are considering the SST field $\mathbf{y}^{\text{SST}} \in \mathbb{R}^{N,T}$ and the global mean net radiative flux at the TOA, $y^{\text{TOA}} \in \mathbb{R}^{T}$. $N$ denotes the number of grid points and $T$ is the length of the simulation. The linear response operator $\mathbf{R}(t)$ is inferred in the piControl run. In order to compute response formulas, we proceed as follows:

\begin{enumerate}[label=(\roman*)]
    \item We run the dimensionality reduction method presented in Section \ref{sec:methods}\ref{sec:dim_red} for the $\mathbf{y}^{\text{SST}}$ field. The identified patterns represent proxies for modes of variability. This step reduces the $N$ dimensional field $\mathbf{y}^{\text{SST}}$ into $n$, \textit{regionally constrained} patterns $c_j$. The identified patterns are shown in Figure \ref{fig:fig_patterns}.

    \begin{figure}[tbhp]
    \centering
    \includegraphics[width=0.5\textwidth]{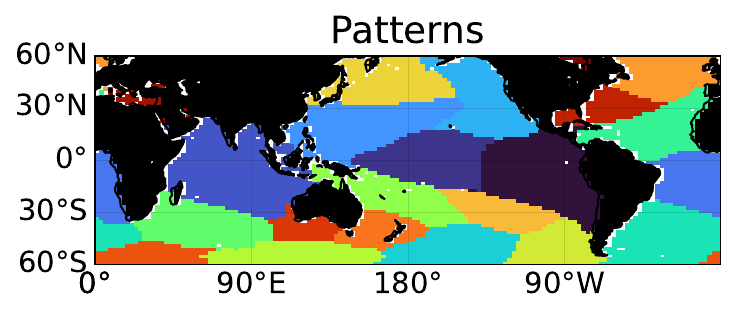}
    \caption{Low-dimensional representation of the sea surface temperature (SST) field. Different colors are used to distinguish different patterns. The observables considered in this study are defined as the integrated SST anomalies in each pattern and by the global mean net radiative flux at the TOA.}
    \label{fig:fig_patterns}
    \end{figure}
    
    \item Given each pattern $c_j$, we compute its SST time series, $X^{\text{SST}}(c_{j},t)$, by computing the integrated SST anomaly inside it as: 

    \begin{equation}
    X^{\text{SST}}(c_{j},t) = \sum_{i \in c_{j}} y^{\text{SST}}_{i}(t) \cos(\theta_{i}),
    \label{eq:signal} 
    \end{equation}
    
    where $\theta_i$ is the latitude at grid point $i$. The inferred time series, $X^{\text{SST}}(c_{j},t)$, as well as the global mean TOA, $y^{\text{TOA}}$, are well approximated by Gaussian distributions (see Appendix C), justifying the use of the approximation presented in Eq. \eqref{eq:response_linear}.
    
    \item The new field $\mathbf{X}^{\text{SST}}\in \mathbb{R}^{n,T}$ and the global mean TOA $y^{\text{TOA}}$ are then concatenated to form the state vector $\mathbf{x} \in \mathbb{R}^{n+1,T}$. At time step $t$, the state of the reduced order system is then encoded in the state vector $\mathbf{x}(t)$. We then compute the regularized covariance matrix $\mathbf{C}_r(0)$ through the regularization procedure and neglect spurious responses at a $\pm 1\sigma$ level, see Section \ref{sec:methods}\ref{sec:numerics_statistics}.

    \item The evaluation step requires us to reconstruct the TOA flux in idealized forced experiments as a function of temperature changes and the response operator $\mathbf{R}(t)$. In the case of the ``coupled'' formulation, we project the SST perturbation, $\delta \mathbf{T}(t)$, in the low-dimensional space in the same way as for $\mathbf{y}^{\text{SST}}$ in point (ii). We then form the perturbation pattern by concatenating $\delta \mathbf{T}(t)$ with a zero forcing pattern in the degree of freedom related to the TOA net flux.
    
    \item Response formulas (Section \ref{sec:methods}) are computed for the low-dimensional system $\mathbf{x} \in \mathbb{R}^{n+1,T}$ using Eq. \eqref{eq:response_convolution} or \eqref{eq:response_step}. The resulting response is referred to as $\delta \langle \mathbf{x}(t) \rangle \in \mathbb{R}^{n+1}$. In this paper we are interested in computing global average responses of the TOA field. To do so, we consider the last entry of response $\delta \langle \mathbf{x}(t)\rangle$, denoted here as $\delta \langle x_{\text{TOA}}(t) \rangle$.
    
\end{enumerate}

We note that the SST and TOA fields are associated with different units and magnitudes. Therefore, the numerical implementation of the protocol benefits from a standardization step, where the response operator and the responses (with Eq. \eqref{eq:response_convolution} or \eqref{eq:response_step}) are computed in a standardized space. Standardization is performed with the standard deviations of the piControl run for each respective variable.

\paragraph*{``Sensitivity map'' metric.} A key metric for understanding the pattern effect is to examine the sensitivity of changes in the global mean net radiative flux at the TOA to local SST perturbations; this is generally approached through ``sensitivity maps". In previous studies, \citep[e.g.,][]{Zhang2023UsingCM4,blochjohnsonetal2024}, sensitivity maps have been estimated by (i) applying a step function perturbation in the SST field at grid point $i$, e.g., a constant $1 K$ for $t > 0$, (ii) computing the change in global mean TOA fluxes after equilibration, (iii) plotting this value at each grid point $i$. These maps show values in units of $[ W / (m^2K) ]$: positive sensitivity corresponds to positive global radiative feedbacks to local SST forcings. Such positive feedbacks decrease the upward radiative response, amplifying the initial global mean temperature changes. The opposite scenario is true for negative sensitivity. In the case of the coarse-grained representation of the system where our variables are not at the grid scale, we define an equivalent metric as follows: given a pattern $c_j$, with $j = 1,...,n$, we prescribe a step function perturbation of 1 Kelvin $[K]$ in each grid point $i$ belonging to $c_j$. The total perturbation in the $j^{\textrm{th}}$ pattern is equal to its area (dimensionality of $[K]$):

\begin{equation}
\Delta T_j = \sum_{i \in c_{j}}  \cos(\theta_{i}) ; ~ \text{for} ~ t>0.
\label{eq:step_function_community} 
\end{equation}

Where $\theta_i$ is the latitude at grid point $i$.  The perturbation field is defined as $\delta \mathbf{T} \in \mathbb{R}^{n+1,T}$, with the $j^{\textrm{th}}$ term equal to $\Delta T_j$ and all other entries equal to zero. We then compute the linear response in global mean net flux $\delta \langle x_{\text{TOA}}(t) \rangle$. The sensitivity map $\mathbf{S}_t \in \mathbb{R}^{N}$ is a gridded map of the same dimensionality $N$ as the original space. The map is defined by plotting at each grid point $i$ belonging to pattern $c_j$, the same value, defined as the global mean TOA response caused by the perturbation $\Delta T_j$:

\begin{equation}
S_{t,i} = \frac{\delta \langle x_{\text{TOA}}(t)\rangle}{\Delta T_j}; ~ \forall i \in c_j,
\label{eq:sensitivity_map}
\end{equation}

The subscript $t$ denotes the time scale up to which the responses are computed. As discussed later in Section \ref{sec:Atmos-Only-Protocol}, studying the ``atmosphere-only'' formulation requires us to focus on the shortest time scales, in this case defined by $t = 1$ month. In this case, the link SST $\rightarrow$ TOA is quasi-instantaneous in agreement with the standard view of feedbacks. In contrast, in the ``atmosphere-ocean coupled'' formulation, we are interested in the equilibrated response to a step function perturbation, obtained by computing the integral in Eq. \eqref{eq:response_step} for $t \rightarrow \infty$. In practice, the equilibrated response can be computed by setting an upper bound of the integral in Eq. \eqref{eq:response_step} that is much larger than the characteristic time of the response; in our case, we take $\tau_\infty = 10$ years, as data are high-passed filtered with a cut-off frequency of $10^{-1}$ years (see Section \ref{sec:data}), so we do not expect any variability beyond some statistical noise at time scales longer than a decade.

\section{Pattern effect in the ``atmosphere-only'' system} \label{sec:Atmos-Only-Protocol}

The traditional formulation of the pattern effect links the change in SST, $\Delta T_i(t)$ at grid point $i$ and time $t$, to the global mean response in TOA flux, $\Delta\overline{\text{TOA}}(t)$, at the same time $t$. The ``Green's function protocol'' \citep{blochjohnsonetal2024} defines a standardized protocol to infer such linkages through perturbation experiments in atmospheric-only models forced by SST boundary conditions. Therefore, the SST field is prescribed and cannot respond to perturbations from either atmospheric or oceanic processes as it would in a true coupled system. In such a setup, the atmospheric model equilibrates to an SST step function perturbation pattern very quickly, and the feedbacks can be considered as instantaneous. More formally, from a data analysis point of view we can consider the feedback as instantaneous when the temporal resolution is coarse enough to be larger than the characteristic time scale of the response. Feedbacks are encoded in $\partial \overline{\text{TOA}}/\partial \text{T}_i$, $\overline{\text{TOA}}$ representing the global mean net flux at the TOA and $\text{T}_i$ the SST at the grid point $i$. We will show that the same approach is contained in our method without the need for expensive model runs.\\

\begin{figure}[tbhp]
\centering
\includegraphics[width=0.5\textwidth]{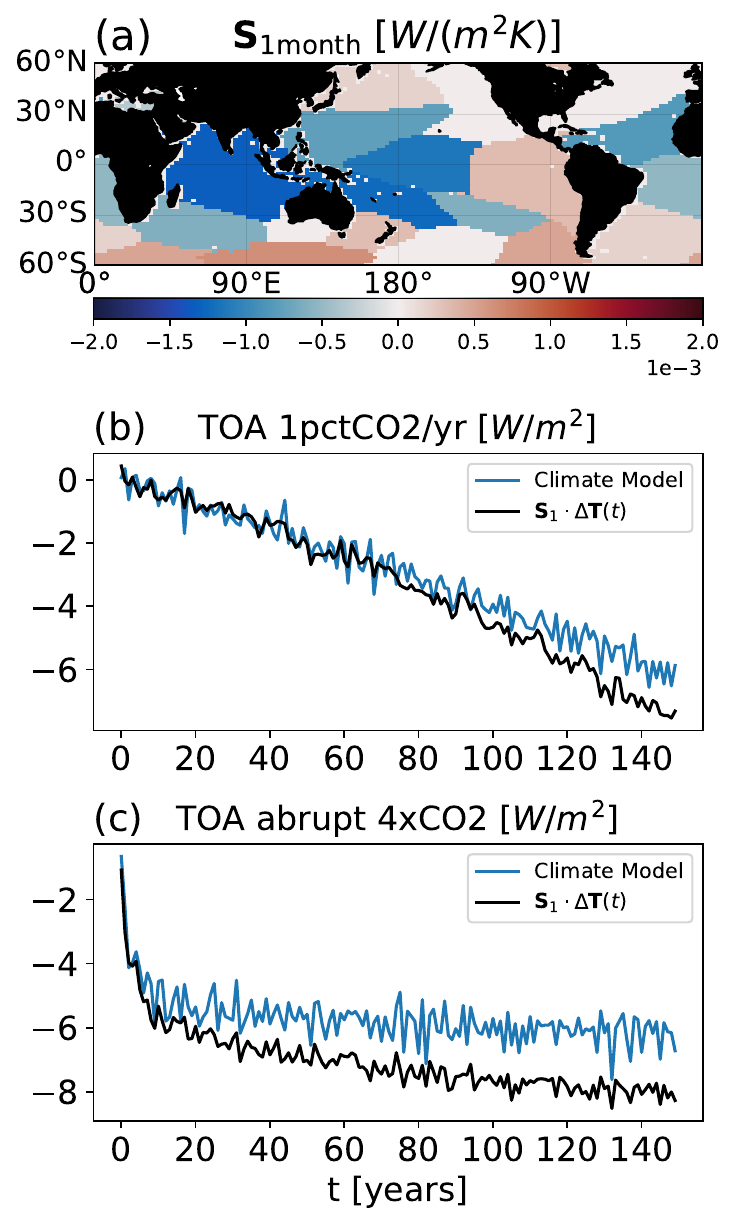}
\caption{Pattern effect in the ``atmosphere-only'' system. Panel (a): Sensitivity maps $\mathbf{S}_{1}$ computed only for the 1 month response (Section \ref{sec:protocol}). Physically, focusing on the shortest time scales allows us to approximate the static sensitivity map in the absence of atmosphere-ocean coupling. At each grid point $i$, we plot the global average response of the net radiative flux at the TOA given a step function SST perturbation of 1 Kelvin imposed at that point. Positive sensitivity corresponds to a positive radiative feedback, therefore amplifying the initial global mean temperature changes; the opposite is true for negative sensitivity. Panel (b,c): Given the sensitivity maps $\mathbf{S}_{1}$ shown in Panel (a) and the sea surface temperature field $\Delta \mathbf{T}(t)$ at time $t$ in the 1pctCO2 and 4xCO2 forced experiments, we compute the dot product $\Delta\overline{\text{TOA}}(t) = \sum_i S_{1,i} \Delta T_i(t)$. Results are shown as yearly averages. This analysis is akin to the one proposed in the ``Green's function protocol'', as in \cite{Zhang2023UsingCM4,blochjohnsonetal2024} and valid for atmosphere-only models where the sea surface temperature is a boundary condition. The response operator has been inferred using the statistical bounds in Section \ref{sec:methods} at the $\pm1\sigma$ level.}
\label{fig:atmos_only}
\end{figure}

We consider the response of the global mean TOA flux to SST perturbations at the shortest time scale $t = 1$ month. Assuming a fast response of the atmosphere compared to the ocean, the subsequent SST change at $t = 1$ month will be small and corresponds to a system without an active atmosphere-ocean coupling such that the SST can be regarded as an imposed boundary condition. This is verified by considering the sensitivity map shown in Figure \ref{fig:atmos_only}(a) computed for $t = 1$ month, which shows strong qualitative similarities to previous approaches involving atmosphere-only models \citep{blochjohnsonetal2024,Zhang2023UsingCM4}.  Specifically, we note the resemblances between our results and \cite{kang2023recent} who used Ridge Regression (compare our Figure \ref{fig:atmos_only}(a) to Extended Data Fig. 7(b) in \cite{kang2023recent}). Our work focuses on seasonal anomalies, and it complements previous work, which focused on annual mean deviations from a global average. A distinct feature of sensitivities at short time scales is a dipole in the tropical Pacific Ocean, with marked negative values on the western side of the basin and positive values on the eastern side, see for example \cite{Dong2019AttributingPacific}. This feature is well captured by our sensitivity map in Figure \ref{fig:atmos_only}(a). The physical mechanisms behind such feedbacks has been described, for example, in \cite{Pascale,williams2023circus}. The positive sensitivity in the eastern Pacific is associated with the high climatology of low-level clouds. A surface warming leads to a decrease in cloud cover, leading to positive anomalies in incoming shortwave radiation. However, negative sensitivity is commonly found in regions of deep convection, such as the western Pacific. An additional warming in these regions travels vertically through the troposphere and horizontally through gravity waves, strengthening the inversion layer in regions with high climatological low-cloud amount. This leads to larger cloud cover in regions such as the eastern Pacific, further increasing the reflected shortwave at the TOA.\\

\begin{figure}[tbhp]
\centering
\includegraphics[width=0.5\textwidth]{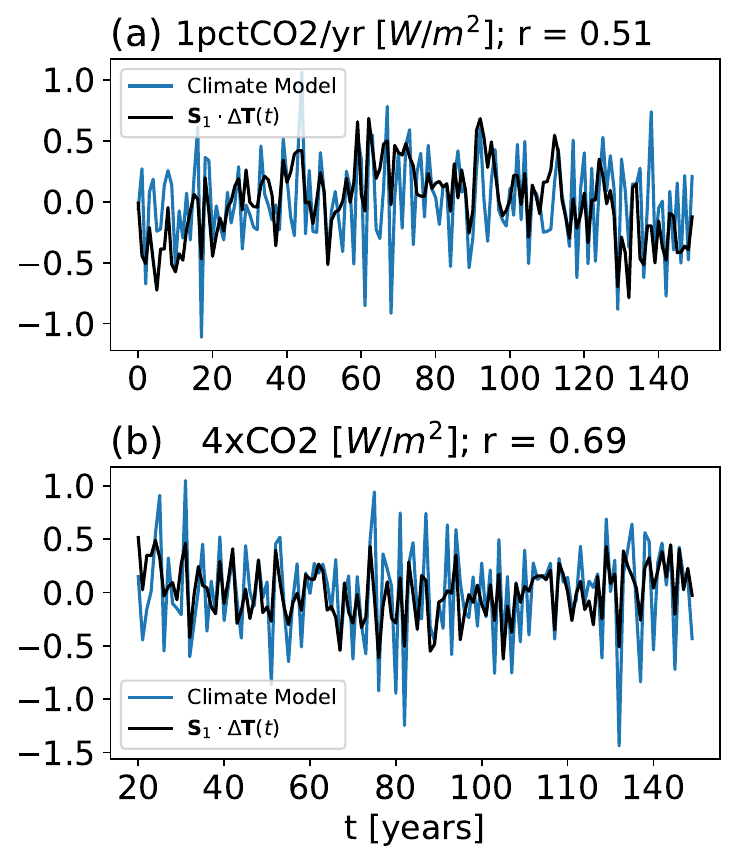}
\caption{Panel (a): Detrended yearly global mean changes in net radiative flux at the TOA as reconstructed by the regression strategy $\mathbf{S}_{1}\Delta \mathbf{T}(t)$ and simulated by the climate model in the 1pctCO2 run. Panel (b): same as panel (a) but for the the 4xCO2 run after removing the first 20 years. The correlation coefficient $r$ between the reconstruction and the model simulation is reported.}
\label{fig:atmos_only-detrended}
\end{figure}

The change in global mean TOA fluxes at time $t$ ($\Delta\overline{\text{TOA}}(t)$) is retrieved as a dot product of the sensitivity map $\mathbf{S}$ with changes in the temperature pattern. This is given by $\Delta\overline{\text{TOA}}(t) = \mathbf{S}_{1}\Delta \mathbf{T}(t) = \sum_i S_{1,i} \Delta T_i(t)$ (each $\Delta T_i(t)$ weighted with the latitudinal weight $\cos \theta_i$), which mirrors Eq. 5 in \cite{blochjohnsonetal2024}. As is standard, we test the framework by reconstructing the change in the global mean net radiative flux at the TOA at time $t$ given the change in the SST field $\Delta \mathbf{T}$ in the 1pctCO2 and 4xCO2 forced experiments introduced in Section \ref{sec:data}). Such reconstructions are computed using monthly anomalies and then shown as yearly averages in Figure \ref{fig:atmos_only}(b,c). The analysis shows a good reconstruction of the global mean TOA flux at each time $t$ given the change in SST at the same time $t$. Quantitatively, for the 1pctCO2 run, the trend in the global mean TOA net flux has been found to be $-0.042 ~ W m^{-2} yr^{-1}$ while the trend in the reconstructed response is $-0.06 ~ W m^{-2} yr^{-1}$. The good skills in reconstructing the TOA response even with such a simple linear method is here ascribed to the coarse-graining procedure. Specifically, given the temporal and spatial resolutions considered, i.e. $1$ month and  $2.5^\circ$ by $2^\circ$, focusing on large areas in the ocean as the patterns in Figure \ref{fig:fig_patterns} allows us to effectively linearize the mapping between changes in temperature $\Delta T_i(t)$ and in global mean TOA fluxes $\Delta\overline{\text{TOA}}(t)$. An additional test for the methodology is to compute the correlation between the reconstruction and the simulated global mean TOA net flux. We remove a trend in both TOA time series and compute their correlation coefficient. The results are shown in Figure \ref{fig:atmos_only-detrended}: The correlation coefficient is relatively well captured by this methodology as $r = 0.51$ and $r = 0.69$ for the reconstruction of 1pctCO2 and 4xCO2, respectively. We note that the analysis in this Section computed the sensitivity maps from the 1 month-lag response operator $\mathbf{R}(1) = \mathbf{C}(1)\mathbf{C}(0)^{-1}$ at a $\pm1\sigma$ confidence level. Importantly, we underscore that no tuning has been used for this experiment and better predictions can be obtained by tuning the size of the regions. However, here we emphasize simplicity and interpretability (i.e. focusing on a few components of the system) rather than more complex models that lead to better evaluation procedures (see also discussion in \cite{HeldGap}).\\

To conclude, the method presented in this section provides a fast and practical tool for studying the pattern effect based on the paradigm of responses to perturbations, akin to approaches in the literature using atmosphere-only models (e.g., \cite{blochjohnsonetal2024,Zhang2023UsingCM4}), but relying solely on a long piControl run of a climate model, avoiding the need for computationally expensive integrations.

\section{Pattern effect in the ``atmosphere-ocean coupled'' system} \label{sec:Coupled-Protocol}

\subsection{Calibration and evaluation} \label{sec:Calibrate}
\begin{figure}[tbhp]
\centering
\includegraphics[width=0.5\textwidth]{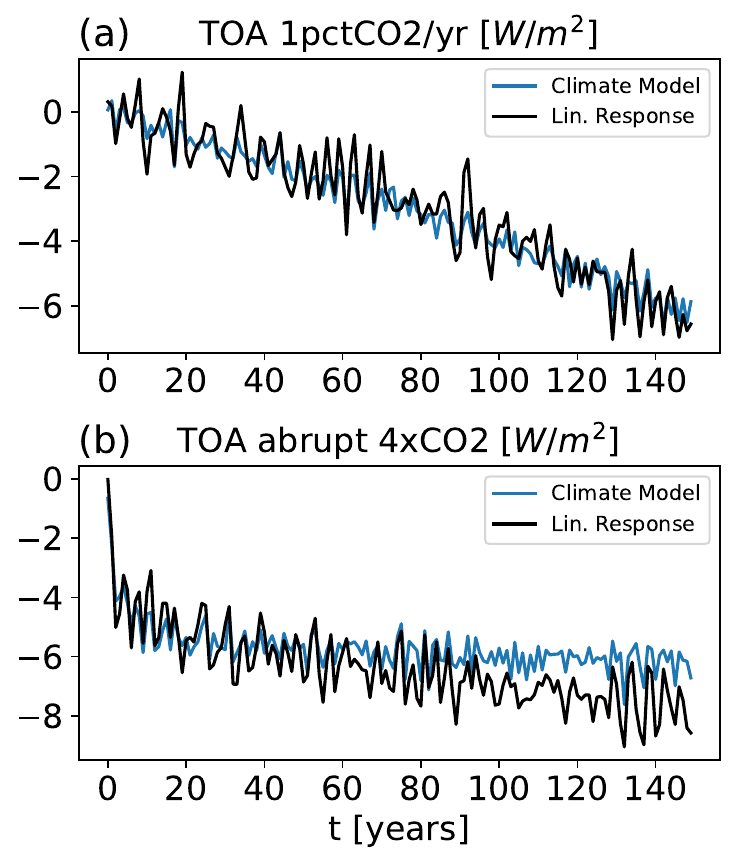}
\caption{Pattern effect in the ``atmospheric-ocean'' coupled system. Panel (a,b): Changes in global mean radiative flux at the top of the atmosphere (TOA) in the 1pctCO2 and 4xCO2 experiments. The net flux at the TOA as simulated by the fully coupled GFDL-CM4 model is shown in blue. A prediction of TOA by linear response theory solely as a function of the changes in the SST field is in black. Linear responses have been computed using the convolution in Eq. \eqref{eq:response_convolution}: the response of TOA at time $t$ is then computed as the integrated effect of the perturbation patterns across all previous time scales $t - \tau$. Results are shown as yearly averages.}
\label{fig:coupled}
\end{figure}

\begin{figure}[tbhp]
\centering
\includegraphics[width=0.5\textwidth]{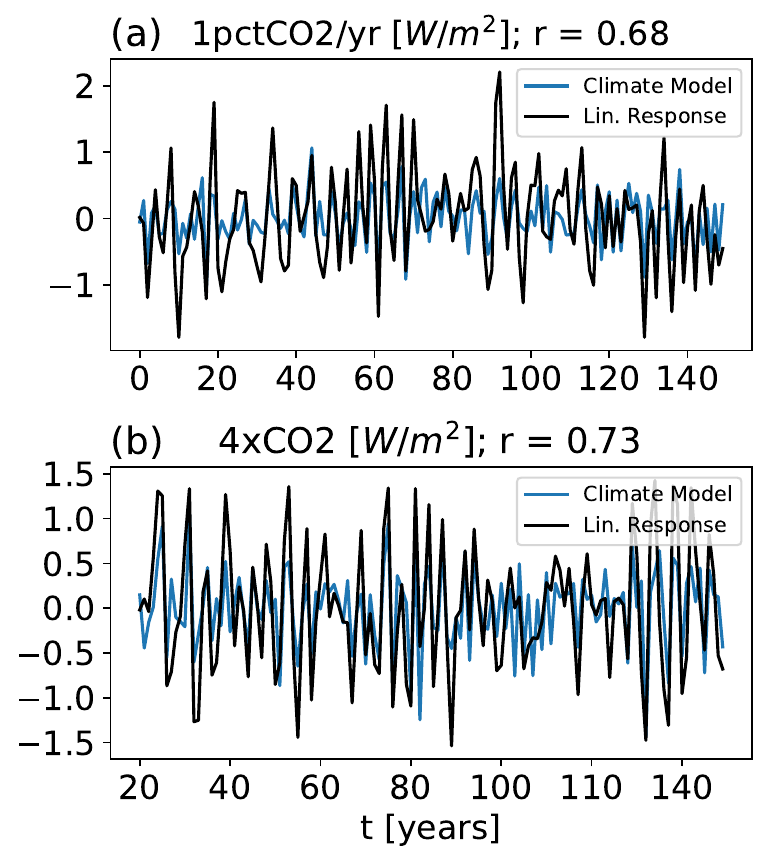}
\caption{Panel (a): Detrended yearly global mean changes in net radiative flux at the TOA as reconstructed by the convolution integral in Eq. \eqref{eq:response_convolution} and simulated by the climate model in the 1pctCO2 run. Panel (b): same as panel (a) but for the the 4xCO2 run after removing the first 20 years. The correlation coefficient $r$ between the reconstruction and the model simulation is reported.}
\label{fig:coupled-detrended}
\end{figure}

In this Section, we consider a new idealized experiment, where we impose step function perturbations in the SST field in the coupled climate system and study the response of the net flux at the TOA. Such a perturbation experiment has not yet been performed with a climate model, but it can be explored within our proposed framework under the assumptions listed in Section \ref{sec:methods} and Appendix A. 
In this case, the SST field is not merely a boundary condition, but an active component of the coupled system: at longer time scales, local SST perturbations can feedback over remote ocean regions through coupled dynamics, e.g. via ``atmospheric bridge''-type mechanisms \citep{Sobel}. Therefore, capturing the appropriate time scales of signal propagation is necessary to accurately build our reduced-order dynamics. Here, the system is viewed through the lens of coarse-grained dynamics, after averaging the SST and TOA flux variables over large regions. However, reducing the spatial dimensionality affects the effective dynamics of the low-dimensional system, which may require a different averaging time scale than the original one-month resolution to satisfy the Markovian assumption. We utilize a practical solution by using the two independent forced experiments, i.e., the 1pctCO2 and 4xCO2 runs. We inferred the response operator $\mathbf{R}(t)$ in the control run after re-processing the data with a reasonable range of temporal resolutions ranging from 3 to 6 months. We tested each implementation against the 4xCO2 simulation and found good agreement for 6-month averages; we therefore chose this temporal resolution. Choosing different temporal resolution mainly affects the magnitude of the mean response, e.g. whether the TOA response after 100 years is $\sim -6 W/m^{2}$ or $\sim -15 W/m^{2}$, but it does not impact the year-to-year variability predictions. We refer to this step as a calibration step. As shown later on, this calibration leads to good performance on the (independent) 1pctCO2 run, thus giving confidence in this data-processing step. Therefore, given the large area of the patterns in Figure \ref{fig:fig_patterns}, the effective 
time scales to capture time-dependent responses to external perturbations is $\sim$ 6 months. In Figure \ref{fig:coupled} we show the reconstruction of the simulated mean radiative flux at TOA in both forced experiments. For the 1pctCO2 run, the model global mean TOA net flux trend is $-0.042 ~ W m^{-2} yr^{-1}$, which is well captured by our reconstruction which is estimated at $-0.043 ~ W m^{-2} yr^{-1}$. As in the previous Section, we compute the correlation of the reconstructed and simulated TOA signals after detrending (Figure \ref{fig:coupled-detrended}). The correlation coefficients are $r = 0.68$ and $r = 0.73$ for the 1pctCO2 and 4xCO2 reconstruction, respectively. As opposed to the atmosphere-only formulation, the variability of the predicted TOA is increased with respect to the model's simulation in this coupled atmosphere-ocean formulation. The 6 months averages for the calibration could be a potential limitation of the framework, as it averages short lags responses.  Considering shorter time scales might be possible by (i) reducing the spatial resolution of the patterns and (ii) adding new state variables relevant at higher frequency and small spatial scales. These considerations are further detailed in Appendix A. 

\subsection{A sensitivity map for the coupled system} \label{sec:Sensitivity_map_coupled}

\begin{figure}[tbhp]
\centering
\includegraphics[width=0.5\textwidth]{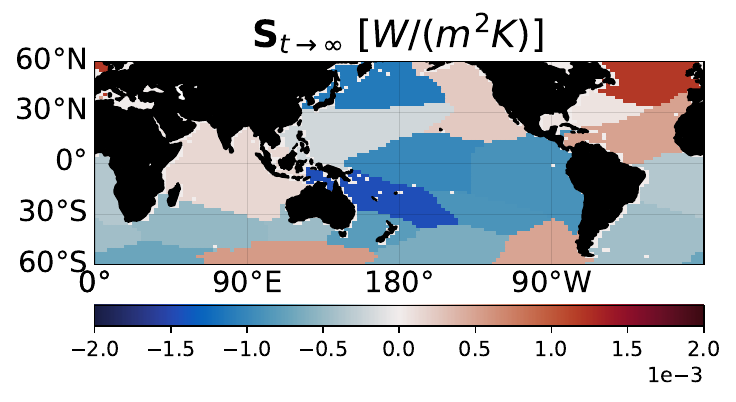}
\caption{Sensitivity map such that at each grid point $i$, we plot the equilibrated, global average response of the net radiative flux at the TOA given a constant perturbation of
1 Kelvin in SST imposed at that point. Positive sensitivity corresponds to a cumulative (in time) positive radiative feedback, therefore amplifying the initial global mean temperature
changes; the opposite is true for negative sensitivity. The time $t \rightarrow \infty$ is here approximated as $t = 10$ years, as all data have been high-pass filtered with a $10^{-1}$ years cut-off frequency.}
\label{fig:coupled-sensitivity-map}
\end{figure}

We now address the following question: What is the cumulative (in time) causal relationship between warming perturbations and changes in the global mean radiative flux at the TOA in a coupled system? In Figure \ref{fig:coupled-sensitivity-map}, we show the equilibrated, cumulative (in time) response of the net TOA flux to a step function perturbation of 1 Kelvin at each grid point in the SST field.  Here, $t = 5$ to 10 years is considered long enough for the atmosphere-ocean coupled system to equilibrate to SST perturbations, see Appendix D. The sensitivity map agrees with previous studies on the large negative sensitivity in the western Pacific but clearly differs in terms of sensitivity of the eastern Pacific, see \citep[e.g.,][]{blochjohnsonetal2024,kang2023recent}. Positive sensitivities (responses) are found in the Indian Ocean, the North Tropical Atlantic, and interestingly, in higher latitudes regions such as the Southern Ocean, the North Pacific, and the North Atlantic. In particular, the North Atlantic carries the largest positive sensitivity, and such sensitivity is found only at longer time scales (around 5 years). However, if the statistical significance is increased to $\pm 3\sigma$, only the tropical domains (and the North Pacific) have non-zero sensitivity, see Appendix E. The positive sensitivity in high-latitudes regions is interesting as it differs to the traditional analysis of the pattern effect but will be left for future studies. We will concentrate mainly on the robust negative sensitivity in the eastern Pacific.\\

The sensitivity map shown in Figure \ref{fig:coupled-sensitivity-map} differs from the one in \ref{fig:atmos_only}(a) by considering longer time scales in the estimation of the response operator and therefore captures the coupled dynamics. 
A temperature perturbation pattern prescribed at the ocean surface impacts the radiative balance at both the TOA and surface on very short time scales. However, over longer periods, an initial warming localized to a specific region can alter dominant modes of variability, redistributing heat globally through coupled dynamics and leading to additional warming and cooling patterns across a broader area than the original perturbation region. In other words, the warming caused by an external perturbation will evolve over time as a result of the coupled dynamics. These cumulative changes can lead to a larger outgoing longwave emission, or short wave feedbacks, than would occur if the warming remained confined to its initial location. Although building a mechanistic understanding of such a cascade of feedbacks is challenging, the response theory formalism integrates these effects across temporal and spatial scales when diagnosing radiative feedbacks \citep{GhilLucarini}. 

\begin{figure*}[tbhp]
\centering
\includegraphics[width=1\textwidth]{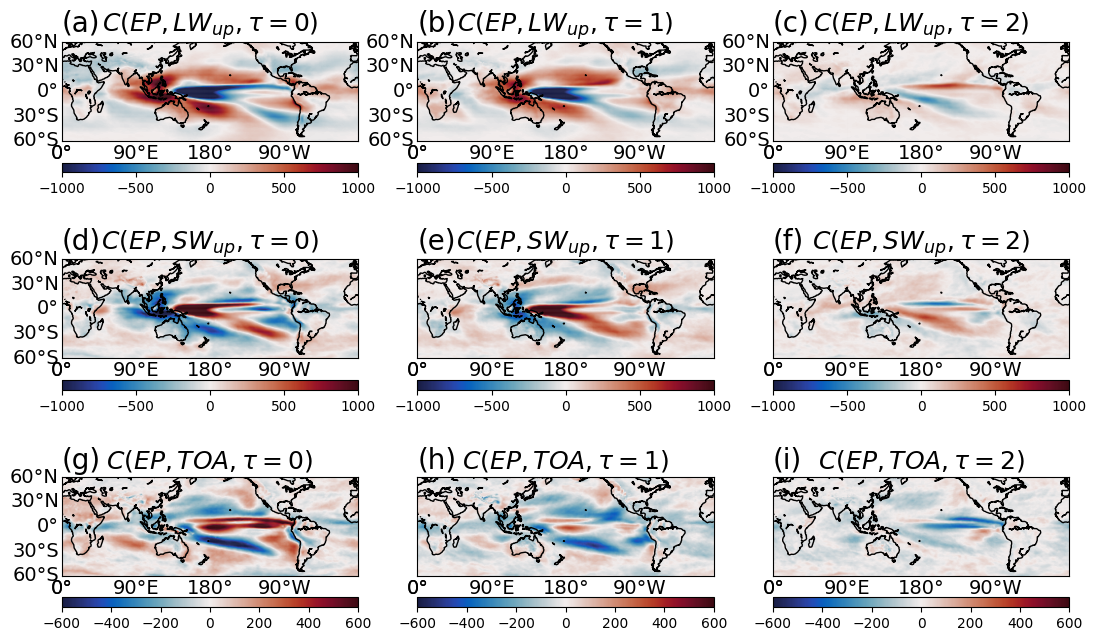}
\caption{In the first row: lag-covariance in $[K W/m^{2}]$ between the eastern Pacific (EP) signal, computed via Eq \eqref{eq:signal} and the outgoing longwave radiation, $LW_{up}$, for $\tau = 0,1,2$. Second and third rows: same as the first row but for the reflected shortwave radiation $SW_{up}$ and the net radiative flux at the TOA, $TOA$. As all quantities are anomalies, the incoming short wave radiation at the TOA is zero and we have $TOA = -SW_{up} -LW_{up}$. The exact following decomposition holds: $C(EP,TOA,\tau) = -C(EP,SW_{up},\tau) - C(EP,LW_{up},\tau)$. Note that the fields are saved at 6-months averages. Negative anomalies of net TOA are seen in the tropical until $\tau = 4$ and fade away after then. Note that the EP signal is computed as the integrated temperature anomaly, not the average, inside the EP region in Figure \ref{fig:fig_patterns}, leading to the large absolute values of covariances seen in the colorbars.}
\label{fig:covariance_enso_6months}
\end{figure*}

\paragraph*{A few insights on the response to eastern Pacific warming.} We further examine changes in radiative fluxes to eastern Pacific warming through a lag-covariances analysis. We stress that covariance does not imply causality, but combined with the causal analysis in the previous section, it can provide additional insights into the time-dependent feedbacks associated with eastern Pacific warming. A constant, external warming applied to the eastern Pacific region drives the system into an El-Ni\~no-dominated state, weakening the temperature gradient across the equatorial Pacific and enhancing deep convection towards the central-to-east Pacific. Additionally, at longer time scales, the persistent local heating over the region forces atmospheric wave responses \citep{Alexander,Sobel} or, at much longer time scales, an oceanic response \citep{BraccoIO,WangThreeOcean}, therefore leading to a heat redistribution over remote ocean basins, changes in cloud cover far from the tropical Pacific Ocean and potentially feedback onto the net radiative flux. To further analyze such changes in net radiative fluxes at the TOA we proceed as follows: (i) we consider the signal (Eq. \eqref{eq:signal}) of the eastern Pacific pattern in Figure \ref{fig:fig_patterns}; (ii) we compute the lag-covariance between the signal and time series of the radiative fluxes at the grid level. We focus on (i) the net flux at the TOA, here referred to as $TOA$ and positive downward (i.e., towards the Earth's surface); (ii) the reflected shortwave radiation, here referred to as $SW_{up}$ and (iii) the outgoing longwave radiation, here referred to as $LW_{up}$. Both $SW_{up}$ and $LW_{up}$ are positive upward (i.e. leaving the planet). As in Section \ref{sec:data}, we are dealing with anomalies, such that the incoming shortwave radiation is zero, and therefore, we have $TOA = - SW_{up} - LW_{up}$. We refer to the covariance between the eastern Pacific signal and radiative fluxes at a lag $\tau$ respectively as $C(EP,TOA,\tau)$, $C(EP,SW_{up},\tau)$ and $C(EP,LW_{up},\tau)$. The following decomposition  holds: $C(EP,TOA,\tau) = - C(EP,SW_{up},\tau) - C(EP,LW_{up},\tau)$. We show the results in Figure \ref{fig:covariance_enso_6months} for three lags, where the lag $\tau = 0$ means synchronous correlations and $\tau = 1,2$ indicates a lead of eastern Pacific towards the radiative fluxes. We remind the reader that the data are saved as 6-months averages. Let us first focus on the $\tau = 0$ analysis. The shift in deep convection from the warm pool region towards the central and east Pacific during an El Ni\~no is captured by the negative covariances with $LW_{up}$ in Figure \ref{fig:covariance_enso_6months}(a). The resulting change in the Walker circulation leads to anomalous sinking in the western Pacific, decreased cloud cover, and, therefore, positive anomalies in  $LW_{up}$. As the eastern Pacific warms, there is a decrease in the stratocumulus deck west of South America, leading to positive anomalies in $LW_{up}$. As the spatial distribution of outgoing longwave radiation changes, so thus the reflected shortwave radiation: an increase in deep convection leads to a decrease in outgoing longwave radiation but also to an increase in reflected shortwave radiation, as shown in Figure \ref{fig:covariance_enso_6months}(d). In general, the changes in reflected shortwave radiation during an El Ni\~no episode are of the opposite sign to the ones in outgoing longwave radiation. The balance between changes in $SW_{up}$ and $LW_{up}$ defines the net radiation change at the TOA and is shown in Figure \ref{fig:covariance_enso_6months}(g). The net change in radiation shows a positive anomaly in incoming radiation in the eastern Pacific and the equatorial Pacific region. Negative anomalies are seen in the north and south of the equatorial Pacific region, as well as in the west Pacific and Indian and tropical Atlantic basins. At lag $\tau = 1$, the covariance between the eastern Pacific signal and the net TOA radiative fluxes are mostly negative, as shown in Figure \ref{fig:covariance_enso_6months}(h). Comparison between Figures \ref{fig:covariance_enso_6months}(b) and \ref{fig:covariance_enso_6months}(e) implies that the negative anomalies in net radiation at longer time scales result from a larger outgoing longwave radiation compared to the incoming shortwave radiation. At longer time lag $\tau = 2$, the main change in net radiation remains negative (Figure \ref{fig:covariance_enso_6months}(i)) for the same reason. Such negative changes intensify up to $\tau = 4$ (not shown), before fading away after. Therefore, the integrated negative changes in net incoming radiative fluxes following a warming of the eastern Pacific can be, in part, explained as a result of a larger cumulative (in time) emitted longwave radiation across the tropical oceans compared to the shortwave radiation feedback.\\

As explained throughout the paper the framework focuses on time-dependent feedbacks that are difficult to address with a covariance analysis only. Consequently, the interpretation above, while useful, remains incomplete. A growing literature has recently investigated the complex, time-dependent linkages among modes of variability in the tropical Pacific, Indian, and Atlantic Oceans and it could serve as future guidance for building mechanistic understandings of the analysis discussed in our work; see, for example, \cite{Pantropical,WangThreeOcean,Capotondi}. To give a practical example, let us consider the observed positive sensitivity in the Indian Ocean (IO) in Figure \ref{fig:coupled-sensitivity-map}. It has been recently recognized that the IO variability can influence both the tropical Pacific and Atlantic through atmospheric and oceanic pathways. For example, a warming projected towards the Indian Ocean Basin mode \citep{Klein} results in enhanced convection over the region. This instability generates eastward-propagating Kelvin waves, which, in turn, can further intensify easterly wind anomalies in the western Pacific, driving the variability in the basin into a La Ni\~na state \citep{Pantropical,WangThreeOcean}. As shown, a warming in the eastern Pacific is associated with a negative sensitivity. It follows that the cooling of the tropical Pacific, as driven by an Indian Ocean warming can lead to a net positive response of the global mean radiative flux at the TOA.

\section{Atmospheric-only and coupled system formulations: conceptual and practical differences} \label{sec:Feedbacks}

We have introduced a comprehensive protocol for studying the effects of idealized warming perturbation experiments from data alone. By examining responses at the shortest time scales or across multiple time scales, these experiments can be framed as atmospheric-only or coupled climate models formulations. Our proposed framework encompasses both approaches, with the primary distinction arising from differences in the formulation of climate feedbacks.

\subsection{Conceptual differences across approaches: climate feedbacks} \label{sec:GF_approach_conceptual_diff}

In the ``Green's function protocol'' \citep{blochjohnsonetal2024, Zhang2023UsingCM4,Dong2019AttributingPacific}, the data-driven sensitivity studies \citep{bloch2020spatial,kang2023recent} and the ``atmosphere-only'' formulation of our protocol (Section \ref{sec:Atmos-Only-Protocol}) radiative feedbacks are encoded in a sensitivity map. In the literature, sensitivity maps are commonly referred to as the radiative feedback and are defined at each grid point $i$ by $\frac{\partial \overline{\text{TOA}}}{\partial \text{T}_i}$, where $\overline{\text{TOA}}$ is the globally averaged net radiative flux at the TOA and $\text{T}_i$ is the SST at point $i$. This perspective is theoretically motivated in the framework of energy balance models by Taylor expanding the radiative response over the steady state temperature (see e.g. Section 2 of \cite{Meyssignac}). The dot product of a sensitivity map with an imposed SST pattern at time $t$ will return an estimate of the globally averaged change in net flux $\overline{\text{TOA}}$ at the same time $t$. This means that an imposed change in SST patterns at time $t$ cannot drive changes in average TOA fluxes at later times $t + \tau$. Naturally, neither our approach (see Section \ref{sec:Atmos-Only-Protocol}) nor the ``Green's function protocol'' \citep{blochjohnsonetal2024} involves a truly instantaneous feedback (i.e., equal to zero at lag $\tau = 0$). Instead, the term ``instantaneous'' refers to the radiative response to imposed SST perturbations occurring on time scales much shorter than the evolution of the SST itself. We note that this is standard in the analysis of climate feedbacks, and is not restricted to the pattern effect \citep{PascaleSW}.\\

The main object of our proposed protocol is the linear response operator, defined as $R_{k,j}(\tau) = \frac{\delta \langle x_k(t+\tau) \rangle}{\delta x_j(t)}$ in the limit of $\delta x_{j}(t) \to 0$. In the general formulation of our framework, where responses are not limited to the shortest time scales, $x_k$ in $\delta \langle x_k(t+\tau) \rangle$ can be a component of the SST or the TOA flux field, responding to a perturbation $\delta x_j(t)$ imposed to the SST or the TOA flux field. Therefore every variable can respond to perturbations in any other variable. Additionally, the operator $R_{k,j}(\tau)$, i.e., the impulse Green's function defined in our framework, is time-dependent as in a physical system, a perturbation in one variable will propagate through the system and impact another variable at later times. The response to external perturbations at time $t$ is then computed as the integrated effect of the perturbation patterns across all previous times $t - \tau$ \citep{Christensen}. In this case, feedbacks are spatially and temporally dependent and encoded in the response operator. This framework allows us to diagnose responses from imposed external perturbations and from the internal time-dependent responses within the coupled system, which further influence radiative feedbacks at later times. The sensitivity map in Figure \ref{fig:coupled-sensitivity-map} then represents the cumulative in time feedback (response) to a perturbation. These different perspectives on feedbacks have been mentioned in the climate literature, see for example both the Introduction and Section 2a in \cite{PascaleSW}. However, response theory provides both a rigorous mathematical foundation and a practical computational tool for their analysis.

\subsection{Practical differences across approaches: next time step prediction vs cumulative effect of perturbations}  \label{sec:GF-Approach-LIM}

The response, using the coupled formulation in its discrete form, is given by 
$\delta \langle x_k(t) \rangle$ to time-dependent perturbations  by $\sum_{j}\sum_{\tau=0}^t R_{k,j}(\tau) \delta x_j(t-\tau) = R_{k,j}(0) \delta x_j(t) + R_{k,j}(1) \delta x_j(t-1) + R_{k,j}(2) \delta x_j(t-2) + ...$. Instead, the standard formulation of feedbacks is encoded in the sensitivity map at the shortest time-scale (of 1 month), namely $\mathbf{S}_{1}$. To evaluate the framework, we focused on reproducing the change in global mean radiative flux at the TOA, as $\Delta\overline{\text{TOA}}(t) = \mathbf{S}_{1}\Delta \mathbf{T}(t) = \sum_i S_{1,i} \Delta T_i(t)$, where $\Delta \mathbf{T}(t)$ are gridded maps of temperature changes. The same approach can be reformulated exactly in the low-dimensional space by considering the first term in the discrete convolution above and studying the change in the net flux at the TOA at time $t$ as $\sum_j R_{k,j}(1) \delta x_j(t-1)$. In this case the variable $x_k$ represents the global mean TOA.\\ 

Thus, there is an equivalence between the traditional evaluation step $\sum_i S_{1,i} \Delta T_i(t)$ and a 1-step prediction $\sum_j R_{\text{TOA},j}(1) \delta x_j(t-1)$. In other words, in our formulation of an ``atmospheric-only'' protocol, there is an exact equivalence between the traditional feedback analysis and a Linear Inverse Model prediction \citep{Penland89}. This further underscores how the conceptual distinctions in the formulation of feedbacks influence the practical implementation of the protocol.\\

Finally, the distinctions in feedback formulations can be well illustrated through the framework of causal inference. Consider a three-dimensional climate system represented as  $x \rightarrow y \rightarrow z$. Here, $x$ and $y$ represent the variability in sea surface temperature in two distinct regions, and $z$ the variability of net radiative fluxes at the TOA. The arrows indicate a causal association: $y\rightarrow z$ implies that a change in temperature in $y$ directly impacts the TOA flux $z$. Following the traditional view of feedbacks, the main radiative feedback will be linked to the direct causal link $y\rightarrow z$. This standard approach is closely aligned with prediction tasks, as demonstrated earlier in this section: information on the variability of $y$ at time $t$ is sufficient to predict $z$ at the next time step. In contrast, the perspective advanced here and in \cite{Lucarini2018} emphasizes the study of how external perturbations propagate along the causal graph. In our simple example, changes in the radiative flux $z$ are directly influenced by SST variability $y$. However, a change in temperature in region $x$ also impacts $y$, subsequently affecting $z$. This second \textit{indirect} link between $x$ and $z$ is captured at longer time scales. This view links feedbacks to the time-dependent ``flow of information'' within a physical system \citep{Ay}, i.e. both direct and indirect causal links, rather than focusing solely on predictions. 

If one's interest is to reconstruct the TOA fluxes then the traditional, atmosphere-only formulation is sufficient, as TOA fluxes respond almost instantaneously (at monthly resolution) to perturbations in SST, leading to the existence of a direct 
causal link between the two variables. The skill in reconstruction is expected to converge across approaches with infinite data and all variables. Our alternative formulation can both perform well for the prediction task and offers novel insights into the complex cascade of coupled feedbacks across spatial and temporal scales, with great relevance for building understanding of causal mechanisms in coupled climate dynamics. Importantly, this distinction highlights how the goals of (i) reconstructing radiative fluxes based on temperature changes and (ii) diagnosing radiative feedbacks do not necessarily have to align. 

\section{Conclusion and discussion} \label{sec:conclusion}

In this work, we developed a protocol based on coarse-graining and the Fluctuation-Response formalism to diagnose and understand the relation between patterns of SST warming and the radiative feedbacks from a long control simulation of a climate model. At least two main classes of methods have been utilized previously to investigate this problem: (i) a Green's function approach, estimating the response of atmospheric fields to local perturbation patches in the ocean using an atmosphere-only model \citep{zhouetal2017,Zhang2023UsingCM4,Dong2020IntermodelModels,Dong2019AttributingPacific, alessi2023surface,blochjohnsonetal2024} and (ii) statistical regression approaches \citep{zhouetal2017,bloch2020spatial,kang2023recent} with coupled climate models. The theory-driven approach presented here, building on the framework recently proposed in \cite{FabCausal}, balances the strengths of each of the previous two methods. Namely, as in previous Green's function methods, it causally studies the atmospheric response to SST perturbations, and it can be applied using only a model's control run, as in previous statistical approaches. Fluctuation-response theory allows us to infer what the response of a dynamical system to small perturbations would have been without actually perturbing the system \citep{Marconi}. The response at time $t$ is then computed by convolving the operator $\mathbf{R}(t)$ with perturbations across all previous time scales $t - \tau$ \citep{Christensen}. 

Our proposed approach is general and simple to apply to study idealized perturbation experiments from data. The method requires two main ingredients: (i) coarse-graining the system (spatially and temporally) in terms of physically relevant, projected dynamics and (ii) computing a response operator utilizing the variability of a long and stationary control simulation. The protocol focused on the dimensionality reduction proposed in \cite{FabCausal}, reducing the system's dimensionality into only a few (here order 10) components. This focus on a small set of patterns facilitates the practical application of FDT in data-scarce scenarios, where only limited variables and samples are available. Comparing with previous studies, \cite{GRITSUN} demonstrated remarkable skills in computing responses with FDT but utilized a dataset comprising four million days at half-day resolution. This extensive dataset included multiple variables and a substantial number of EOFs (1800), allowing perturbations to be projected near the grid scale. Our approach emphasizes working with significantly fewer samples and only two variables, reflecting the constraints of real-world scenarios with limited data availability. This constraint necessitates a coarse-grained, stochastic representation of the system's variability, achieved by averaging over large spatial regions and time scales, with variables active at finer spatial and temporal scales modeled as noise. Importantly, such choices are constrained by various assumptions about the system's dynamics, and we further discuss their limitations and strengths in Appendix A. In general, no single dimensionality reduction tool is expected to be optimal for all applications, and a key area for future research will involve exploring various methods to decompose the original datasets, tailored to the available sample size and the specific type of dynamics investigated.\\


Traditional experiments that diagnose the response of atmosphere-only models to SST boundary conditions can be reframed within the proposed framework by focusing on the shortest time scales, under the assumption that the response at short time scales is dominated by the atmospheric dynamics. When longer time scales are considered, the protocol enables idealized climate experiments that capture the joint evolution of the atmosphere and the slow oceanic dynamics. The long term memory of the sea surface temperature field \citep{KlausF,Galfi} is the key physical factor enabling differentiation across the two frameworks. In the context of the pattern effect, our framework reproduces results consistent with existing literature in the atmosphere-only formulation. However, when the coupled dynamics is included -- accounting for cumulative responses over extended time scales -- the framework generates a novel sensitivity map, offering a qualitative prediction of TOA radiative flux responses to SST perturbations in coupled climate model experiments. The sensitivity map in Figure \ref{fig:coupled-sensitivity-map}, reveals a negative response to SST warming throughout the tropical Pacific. This finding contrasts with instantaneous feedbacks reported in the literature (and in our ``atmosphere-only'' implementation), which emphasize a dipole sensitivity pattern in the tropical Pacific, with positive and negative sensitivities on the east and west sides of the basin, respectively. In both cases, i.e. ``atmosphere-only'' and ``coupled'' formulation, the framework allows us to perform well at the ``prediction" task and reproduce changes in the net radiative flux in forced simulations. The key difference between the two formulations is an additional perspective on the theory of climate feedbacks. The traditional approach defines climate feedbacks in terms of instantaneous relationships and it is centered around the problem of prediction/reconstruction. Instead, the ``coupled' formulation of our framework extends this concept to account for spatially and temporally dependent linkages across climate fields. This broader perspective aligns with the theoretical proposals of \cite{Lucarini2018} and focuses on understanding how perturbations propagate through the coupled climate system.\\

The ``coupled'' formulation presents an intrinsically more challenging problem than the traditional analysis of pattern effect. In this case -- diagnosing how perturbations propagate throughout the system across spatial and temporal scales -- focusing on only two variables represents a more significant simplification than in the ``atmosphere-only'' formulation. Future studies should aim to explore the ``coupled'' formulation across a broader set of variables and finer spatial and temporal scales. Nonetheless, we consider this study a valuable starting point for such future exploration and a preliminary blueprint for future perturbation experiments with coupled climate models.\\

A clear limitation of this work is that we focused exclusively on the GFDL climate model. As is well-known, climate models are affected by model errors and they are limited in their ability to represent the ``real'' climate \citep{LennyPNAS}. Future work should aim to move beyond model land \citep{ThompsonSmith} and explore the applicability of the proposed framework to real-world observations. A significant constraint in this direction is the extremely limited time span of reliable satellite observations on radiative fluxes, which began only in 1998 \citep{ceres}. A potential solution could involve integrating the proposed framework with Bayesian analysis, where ``prior'' response functions are inferred from a catalog of climate models, under the assumption of a reduced model error through multi-model ensemble analysis. This prior response functions could then be updated using a substantial portion of observational data and tested on the remainder. Furthermore, future research could investigate the pattern effect from the theoretical perspective introduced by \cite{RUELLE1998220,RuelleReview} and its computational implementation in \cite{Lucarini2017,Lembo}. This would allow for a deeper quantification of the limitations/strengths of the assumptions and preprocessing procedures proposed in this work.\\

Finally, the results in this work serve as additional evidence for the relevance of FDT in climate studies, even in its simple quasi-Gaussian approximation, after carefully choosing the relevant observables. Given appropriate consideration of the coarse-graining steps, the FDT can be utilized throughout climate science beyond what is examined here, with great relevance for inferring causal linkages and building understanding of climate dynamics.


\appendix[A]

\appendixtitle{A few considerations on the application of linear response theory for spatiotemporal climate data} \label{sec:note_on_FDT}

We briefly review current applications of linear response theory in climate science, including the limitations of the fluctuation-dissipation Theorem (FDT). We then highlight important considerations for applying the fluctuation-response formalism in the climate context. These considerations are important steps for the effective application of FDT to climate data.

\subsection{A summary of linear response theory in climate science}

The Earth's climate is a complex, spatiotemporal dynamical system with variability across a large range of spatial and temporal scales \citep{GhilLucarini}. It has been recently argued that linear response theory serves as a comprehensive framework to understand and quantify (i) large scale climate dynamics and (ii) its response to external forcings. As outlined in the main text, there has been two main versions of linear response theory applied to climate problems: the Fluctuation-Dissipation Theorem/Relation \citep{Leith, Marconi,MajdaBook} and Ruelle response theory \citep{RUELLE1998220,Lucarini2017,LucariniChekroun}. In the last two decades there has been active research on strengths and limitations of both approaches \citep[see e.g.,][]{LucariniReviewGeophysics,GritsunValerio,Christensen,LucariniChekroun}.\\ 

\cite{RUELLE1998220} proposed a new perspective on linear response grounded in dynamical system theory rather than near-equilibrium statistical mechanics as the original formulation of FDT \citep{Kubo,Sarracino}. This different perspective is recently emerging as a general, rigorous tool to study and attribute changes in the climate system to external forcing with impressive results at both global and regional levels \citep[e.g.,][]{Lucarini2017,Lembo}. In practice, the general strategy is to define a Green's function through a few simulations of a climate model, for example, by using a control and a step-function run. It is then possible to convolve the Green's function with, for example, new $\text{CO}_2$ forcing and investigate different possible climate change scenarios \citep[e.g.,][]{Lembo}. Recently, \cite{LucariniSantos} showed how to link the forcing to free modes of variability in the context of Ruelle linear response, therefore adding to interpretability and understanding of the system's response. As with FDT \citep{Aurell,Baldovin}, Ruelle response theory is causal, in the sense of interventional causality \citep{IsmaelNew,Pearl2008}. We highlight the contributions by \cite{LucariniColangeli,Lucarini2018,Valerio2021,Lembo,Basinski,LucariniChekroun} within the broader framework of Ruelle response theory. In particular, our work shares strong similarities with the proposal of \cite{Lucarini2018}. Such approaches are highly relevant for future studies on the pattern effect and, more generally, as rigorous methods for studying feedbacks and performing causal attribution of climate change \citep{Valerio2024}.\\

The FDT formalism, as considered in this study, is less general than the strategy outlined above \citep{GhilLucarini}, and it has been argued that results can be affected by dimensionality reduction procedures \citep{Pedram2}, by the variables of choice \citep{GritsunValerio}, by the length of the dataset analyzed \citep{LucariniReviewGeophysics} and on whether the forced response in question projects strongly onto the internal variability \citep{GritsunValerio}.

\subsection{Practical application of FDT: assumptions, limitations and strengths}\label{sec:note_on_FDT_sub}

Despite drawbacks, the FDT has proven to be relevant and useful in climate studies \citep[e.g.,][]{MajdaBook,GRITSUN,Lacorata,Haynes} and, in general, in dynamical systems with many degrees of freedom \citep{Colangeli2012,Sarracino}. In its domain of applicability, the FDT approach is extremely powerful as it eliminates the need to perform new simulations to construct Green's functions and focus only on long stationary simulations or, ideally, on observational data.\\

While the theoretical formalism behind the FDT framework is general, its practical implementation is often non-trivial and influenced by coarse-graining procedures, data preprocessing, and, most critically, the underlying assumptions about the system. We highlight several important limitations and caveats - often overlooked in the literature - that may affect interpretations in future studies. The FDT formalism enables us to study how external perturbations propagate in a dynamical system, assuming access to the full state vector $\mathbf{x}(t) = [x_{1}(t),x_{2}(t),...,x_{N}(t)]$. However, in practice, this is rarely the case, necessitating coarse-graining procedures and a stochastic reformulation of the system. In our analysis, this led to specific choices: (i) averaging sea surface temperature (SST) variability over large spatial regions, (ii) using globally averaged net radiative flux at the TOA, and (iii) adopting a temporal resolution of six-month averages. These decisions were guided by a calibration process (see Section \ref{sec:Coupled-Protocol}\ref{sec:Calibrate}) and prior studies. For instance, earlier work demonstrated that coarse-grained sea surface temperature (SST) fields, particularly in the tropics, can be modeled as a Markov process \citep{Penland89,Penland95}, while more recent research has identified SST and top-of-atmosphere (TOA) net radiative flux as key variables for studying the pattern effect \citep{blochjohnsonetal2024,Dong2019AttributingPacific}. However, the climate system encompasses far more variables than just SST and TOA radiative flux. In this context, coarse-graining becomes essential, treating processes at smaller spatial and faster temporal scales as noise. Crucially, such methodological choices -- common to all applications of the FDT -- are not dictated by rigorous mathematical criteria but are based on assumptions made by researchers. These assumptions should be taken into account as they significantly shape the results and their interpretation. We also briefly note that dealing with incomplete observations is a common problem across many fields. In order to reconstruct the full system, there have been proposals to consider applications of the Takens' embedding theorem \citep{Takens}. The practical implementation of such strategy is, however, severely constrained by the dimensionality of the system and it is not valid for stochastic systems. Therefore, Takens' theorem cannot be a valid option in this study and we refer to the discussions in \cite{Baldovin,Lucente} for more details on this point.\\

The FDT application in this work focused on a simple form of FDT presented in Eq. \eqref{eq:response_linear}, referred to as ``quasi-Gaussian approximation'' by \cite{MajdaBook}. This form of FDT is the one used in many previous applications \citep[e.g.,][and references therein]{Pedram2} and it is valid for linear systems. The climate system is nonlinear and it is therefore not obvious why Eq. \eqref{eq:response_linear} should work. Again, the coarse-graining procedures play a central role in enabling the use of this simplified form. As noted, in several previous works, \citep{Prashant,MajdaStructuralStability}, the probability distribution of coarse-grained climate variables is often smooth and Gaussian. This holds true in our analysis as well, as shown in Appendix C. Thus, the use of Eq. \eqref{eq:response_linear} is justified.\\ 

Finally, we note that the name ``quasi-Gaussian'' rather than ``Gaussian'' approximation, stresses the fact that while $\rho(\mathbf{x})$ is approximated as a Gaussian, lag-covariances are computed at each time $t$ by averaging over the data (whether with linear or nonlinear dynamics), rather than automatically assuming linear dynamics with Gaussian invariant density. This is an important difference from linear regression modeling strategies, leading for example to skill in capturing changes in variance in contrast to linear inverse models (see \cite{Majda2010} for details and in-depth comparisons). 

\appendix[B]

\appendixtitle{Dimensionality reduction through community detection}

For completeness we report here the main steps of the dimensionality reduction step proposed in \cite{FabCausal}, and refer to that paper for further details. Consider a spatiotemporal field saved as a data matrix $\mathbf{x} \in \mathbb{R}^{N,T}$. $N$ is the number of grid points and $T$ is the length of each time series. For example, $\mathbf{x}$ could be the sea surface temperature field. The dimensionality reduction proposed in \cite{FabCausal} works in a few simple steps:

\begin{itemize}
    \item Compute the correlation matrix $\mathbf{C}$, defined as $C_{i,j} = \overline{x_{i}(t) x_{j}(t)}$, where the overline stands for temporal averages, and each $x_{i}$ has been scaled to zero mean.
    \item Define an Adjacency matrix $\mathbf{A}$ from the correlation matrix $\mathbf{C}$ by setting $A_{i,j} = 1$ if (i) $C_{i,j}$ exceeds a threshold $k$ and (ii) the distance between grid points $i$ and $j$ is smaller than a threshold $\eta$. If (i) and (ii) are not satisfied, then $A_{i,j} = 0$. $\mathbf{A}$ is the matrix representation of a graph where nodes $i$ and $j$ are connected if sharing large covariability and if they are close on a longitude-latitude grid. Regionally constrained patterns of variability can then be identified by finding ``communities'' in the graph \citep{Barabasi,Newman,Lancichinetti}. With communities of a graph, we refer to group of nodes that are much more connected to each other than to the rest of the graph. Importantly, parameters $k$ and $\eta$ are automatically defined by two simple heuristics. The two heuristics depend on two parameters $q_k = 0.95$ and $q_\eta = 0.1$ and we refer the reader to \cite{FabCausal} for details. We chose a value of $q_\eta = 0.1$ rather than $0.15$ as in \cite{FabCausal} in order to split the ENSO region (Figure 2 in \cite{FabCausal}) into an eastern and central Pacific region.
    \item Each node $i$, correspondent to a grid point $i$ on the map, will then be associated to a community/pattern. In other words, we partitioned a spatiotemporal climate field of spatial dimension $N$, in a series of $n$ regions $c_j$, with $j = 1,...,n$. We identify communities through the Infomap community detection algorithm \citep{Rosvall1,Rosvall2,MapEq,RosvallReview} as shown in \cite{FabCausal}. The size and number of the identified patterns will roughly depend on the $q_k$ and $q_\eta$ parameters presented above.
    \item Finally, to each community $c_j$, we are going to associate a time series defined as the integrated anomaly inside, i.e. $X(c_{j},t) = \sum_{i \in c_{j}} x_{i}(t) \cos(\theta_{i})$. Where 
    $\theta_{i}$ represents the latitude at grid point $i$ and $\cos(\theta_{i})$ a latitudinal scaling.
\end{itemize}

To summarize, given a spatiotemporal field saved as a data matrix $\mathbf{x} \in \mathbb{R}^{N,T}$, the proposed framework allows us to define a new field $\mathbf{X} \in \mathbb{R}^{n,T}$, with $n \ll N$.

\appendix[C] 

\appendixtitle{Probability distributions} \label{sec:PFDs}

\begin{figure}[tbhp]
\centering
\includegraphics[width=0.5\textwidth]{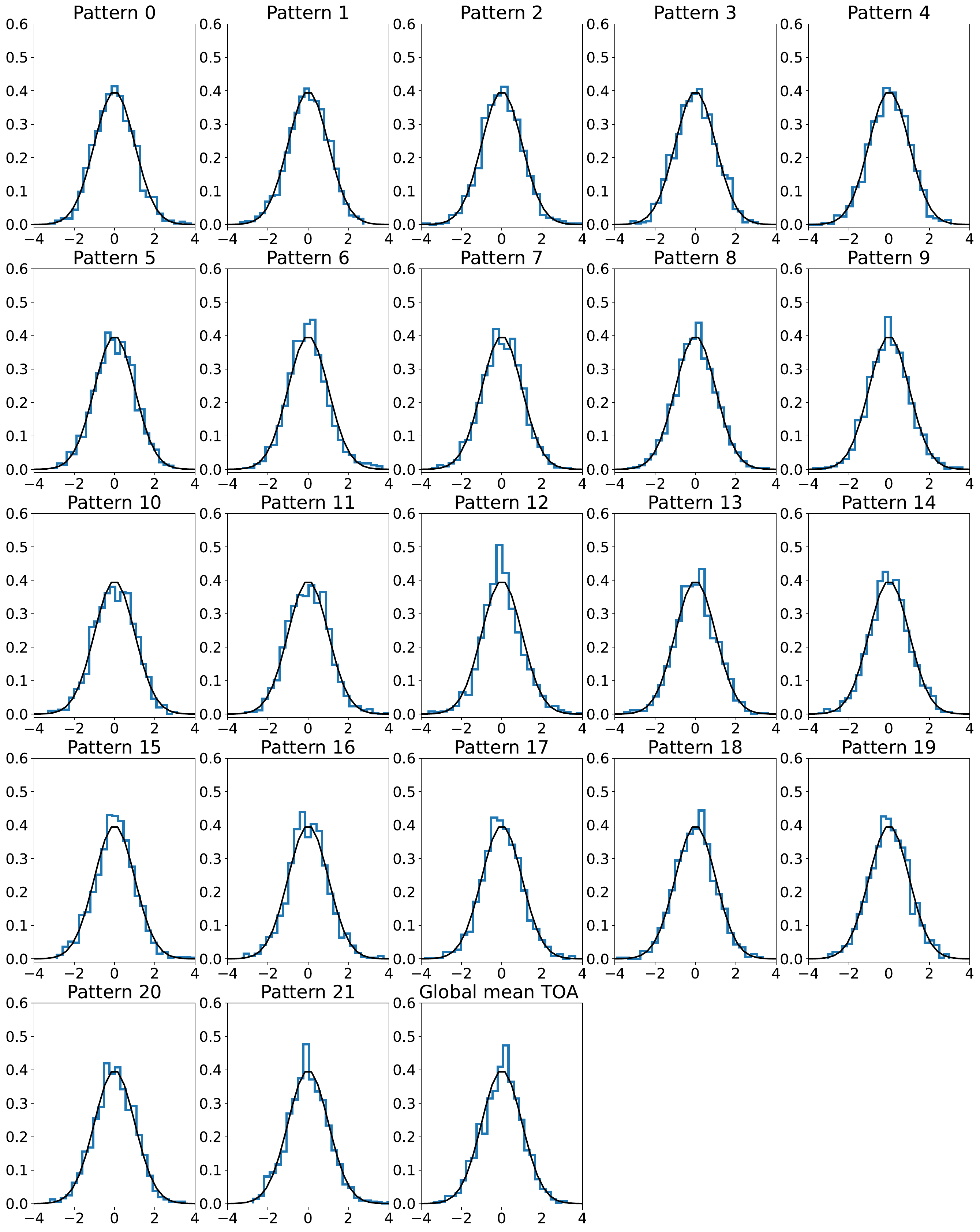}
\caption{Probability distributions of the cumulative time series of sea surface temperature in each pattern in Figure \ref{fig:fig_patterns} and of the global mean net radiative flux at the TOA. Each signal had the mean removed and has been standardized to unit variance. A Gaussian fit with zero mean and unit variance is shown in black on top of each histogram.}
\label{fig:fig_1C}
\end{figure}

In Figure \ref{fig:fig_1C}, we show the histogram of the integrated SST anomalies in each one of the patterns in Figure \ref{fig:fig_patterns} and of the global mean net radiative flux at the TOA. Each time series have been scaled to zero mean and unit variance. A standard normal distribution is also shown in black for comparison. This analysis demonstrates that the quasi-Gaussian approximation shown in Eq. \eqref{eq:response_linear} is indeed relevant for the system studied. The Gaussianity of the process is a direct consequence of our preprocessing by coarse-graining in both the temporal and spatial directions, further confirming the ideas and findings of previous papers such as \cite{Prashant}.

\appendix[D] 

\appendixtitle{Diagnosing the characteristic time scale of the response} \label{sec:convergence}

\begin{figure}[tbhp]
\centering
\includegraphics[width=0.5\textwidth]{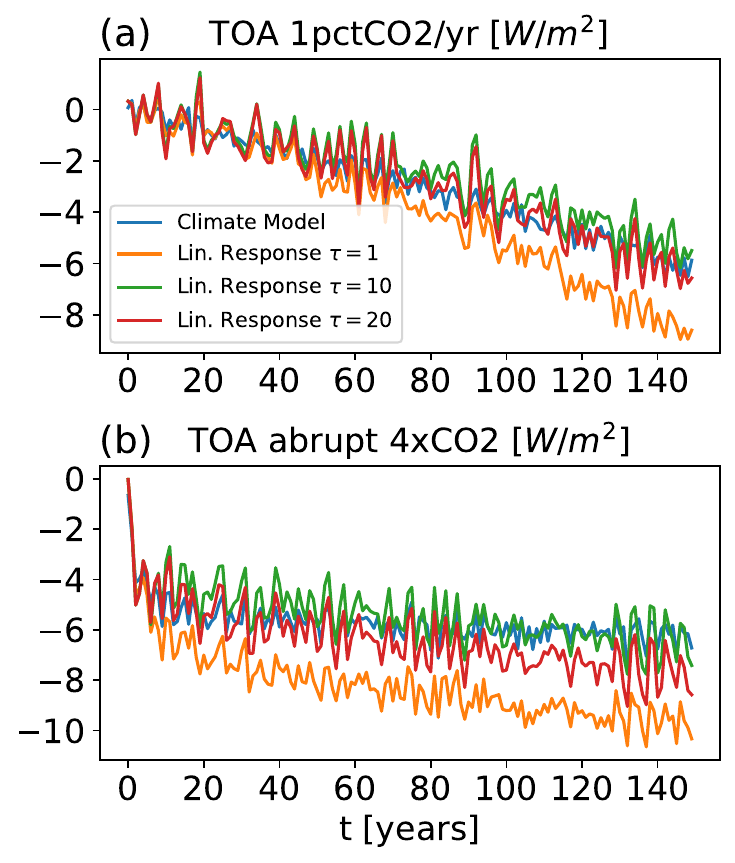}
\caption{Pedicting the change in the net radiative flux at the top of the atmosphere (TOA) as a function of a time-scale parameter $\tau$. We compute responses of the global mean TOA to changes in SST in the forced simulation using after setting $R_{k,j}(t) = 0$ for $t \geq \tau$, with $\tau = 1,10,20$, correspondent to + 6 months, + 5 years and + 10 years.}
\label{fig:fig_1D}
\end{figure}

The sensitivity map in Figure \ref{sec:Coupled-Protocol} integrates all responses up to $t = 10$ years. Such threshold has been considered mainly because data have been high-pass filtered with a cut-off frequency of $10^{-1}$ years. Therefore variability present for $t \geq 10$ years is here considered as noise. However, the characteristic time scale of the response in TOA can appear sooner than 10 years. In fact, the response operator is defined as the response to a small impulse perturbation. Therefore, at long time scales ($t \to \infty$), the response $R_{k,j}(t)$ between any variable $x_j$ and $x_k$ should (i) go to zero, see for example Figure 1 in \citep{Baldovin}, or (ii) become statistically insignificant, see for example Figure 1 in \citep{FabCausal}. In practice, then, it is common not to compute the response operator $\mathbf{R}(t)$ for all $t$ and set $R_{k,j}(t) = 0$ after a long characteristic time scale $\tau$, see for example \cite{Pedram2}. The value of $\tau$ will depend on the system itself. To explore this time scale (and the robustness of our choice of 10 years) we then evaluate the framework as done in the main text (i.e. by reconstructing the change in TOA mean flux given changes in the Temperature field) while setting to zero every $R_{k,j}(t)$ for $t>\tau$, with $\tau = 6$ months, 5 and 10 years. Results are shown in Figure \ref{fig:fig_1D}. Independent on the tested data, i.e. 1pctCO2 ot 4xCO2 experiment, considering the impact of longer time scales, helps reducing the bias in the reconstruction (i.e. $\tau = 5$ and 10 lead to better reconstruction of $t = 1$). Small differences are observed when going from $\tau = 5$ to 10 years in the 1pctCO2, with slightly better values observed for $\tau = 10$ years. In the case of 4xCO2, both $\tau = 5$ and $\tau = 10$ years result in relatively good reconstructions. For $\tau = 5$ years, better reconstructions are observed after approximately 70 years compared to $\tau = 10$ years, but the reconstruction is worse in the first 60 to 70 years. We conclude that a time scale of $\tau \geq$ is long enough to observe the cumulative response to perturbations. Note that the confidence bounds cosidered here can further influence this analysis, and we would expect to have clearer results in the case of much longer datasets.

\appendix[E] 

\appendixtitle{A sensitivity map in the coupled system using strict confidence bounds} \label{sec:sens_map_3sigma}

We plot the sensitivity map as in Section \ref{sec:Coupled-Protocol} but focusing on a very strict confidence bound of $\pm 3\sigma$. In this case only tropical domains becomes relevant to describe the pattern effect. However, such strict confidence bounds almost surely mask also ``true'' responses and future work should focus on analysis on much larger datasets or on multi-model ensembles.

\begin{figure}[tbhp]
\centering
\includegraphics[width=0.5\textwidth]{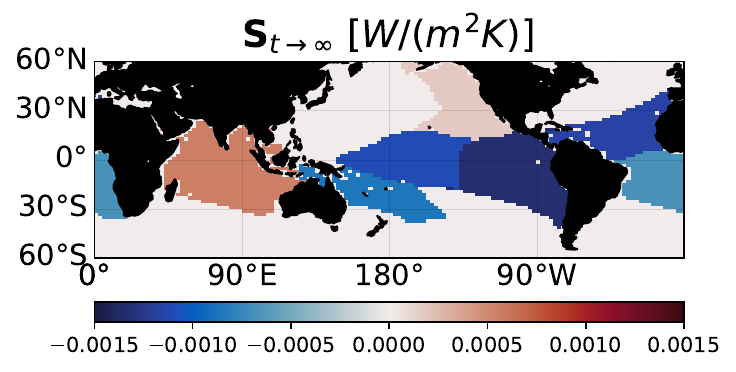}
\caption{Sensitivity map such that at each grid point $i$, we plot the equilibrated, global average response of the net radiative flux at the TOA given a constant perturbation of
1 Kelvin in SST imposed at that point. Positive sensitivity corresponds to a cumulative (in time) positive radiative feedback, therefore amplifying the initial global mean temperature
changes; the opposite is true for negative sensitivity. The time $t \rightarrow \infty$ is here approximated as $t = 10$ years, as all data have been high-pass filtered with a $10^{-1}$ years cut-off frequency.}
\label{fig:fig_1E}
\end{figure}

\clearpage
\acknowledgments
The authors would like to thank two anonymous reviewers and Valerio Lucarini for their comments and suggestions which greatly improved the quality of the manuscript. F.F. acknowledges numerous discussions with Pavel Perezhogin. This work was supported by  the NOAA Grant  NA20OAR4310411, National Science Foundation Grant  OCE‐2048576, and  by the KITP Program “Machine Learning and the Physics of Climate” supported by the National Science Foundation under Grant No. NSF PHY-1748958.

%
%
\datastatement
Codes for the community detection and implementation of Fluctuation-Dissipation formulas, can be found in https://github.com/FabriFalasca/Linear-Response-and-Causal-Inference. 


\bibliographystyle{ametsocV6}
\bibliography{references}

\end{document}